# A Ferroelectric Semiconductor Field-Effect Transistor

Mengwei Si[1,4], Atanu K. Saha[1], Shengjie Gao[2,4], Gang Qiu[1,4], Jingkai Qin[1,4], Yuqin Duan[1,4], Jie Jian[3], Chang Niu[1,4], Haiyan Wang[3], Wenzhuo Wu[2,4], Sumeet K. Gupta[1], and Peide D. Ye[1,4],*

[1]School of Electrical and Computer Engineering, Purdue University, West Lafayette, Indiana 47907, United States

[2]School of Industrial Engineering, Purdue University, West Lafayette, Indiana 47907, United States

[3]School of Materials Science and Engineering, Purdue University, West Lafayette, In 47907, United States

[4]Birck Nanotechnology Center, Purdue University, West Lafayette, Indiana 47907, United States

*Address correspondence to: yep@purdue.edu (P.D.Y.)

**Abstract**

Ferroelectric field-effect transistors employ a ferroelectric material as a gate insulator, the polarization state of which can be detected using the channel conductance of the device. As a result, the devices are of potential to use in non-volatile memory technology, but suffer from short retention times, which limits their wider application. Here we report a ferroelectric semiconductor field-effect transistor in which a two-dimensional ferroelectric semiconductor, indium selenide (α-$In_2Se_3$), is used as the channel material in the device. α-$In_2Se_3$ was chosen due to its appropriate bandgap, room temperature ferroelectricity, ability to maintain ferroelectricity down to a few atomic layers, and potential for large-area growth. A passivation method based on the atomic-layer deposition of aluminum oxide ($Al_2O_3$) was developed to protect and enhance the performance of the transistors. With 15-nm-thick hafnium oxide ($HfO_2$) as a scaled gate dielectric, the resulting devices offer high performance with a large memory window, a high on/off ratio of over $10^8$, a maximum on-current of 862 μA μm$^{-1}$, and a low supply voltage.

Ferroelectric (FE) materials exhibit a spontaneous polarization in the absence of an external electric field. This polarization can be reoriented by ion displacement in the crystal and polarization switching can be triggered by an external electrical field such that ferroelectric materials can have two electrically-controllable non-volatile states[1]. As a result, ferroelectric random access memory (FeRAM) has long been studied as a non-volatile memory technology[2–14]. FeRAM uses a ferroelectric capacitor to build a one-transistor–one-capacitor (1T1C) cell. However, the reading process in the capacitor type FeRAM is destructive and requires rewrite after each reading operation. This structure has been commercialized but with a limited market share.

Alternatively, ferroelectric field-effect transistors (Fe-FETs; Fig. 1a) can be used to build a type of one-transistor (1T) non-volatile memory. In a Fe-FET, a ferroelectric insulator is employed as the gate insulator in a metal–oxide–semiconductor field-effect transistor (MOSFET). The channel conductance is used to detect the polarization state in the ferroelectric gate insulator so that the data reading operation in Fe-FETs is non-destructive. Fe-FET is a promising memory technology due to the fast switching speed in ferroelectric materials (nanoseconds or less[4,9,11,15,16]), its non-destructive readout, its non-volatile memory state and its simple structure for high density integration.

This Fe-FET structure was, however, first proposed in 1957 (ref. 17) and has not yet been commercialized because of its short retention time. The two major causes of the short retention time are the depolarization field and the gate leakage current. The depolarization field is the result of the potential drop across the interfacial dielectric and the band bending of the semiconductor, which leads to charge trapping at the ferroelectric insulator/semiconductor interface[2,6,14,18]. Therefore, charge trapping and gate leakage current can cause charge accumulation at the ferroelectric insulator/semiconductor interface, which leads to threshold voltage ($V_T$) drift and the destruction of the memory state.

In this Article, we report a ferroelectric semiconductor field-effect transistor (FeS-FET), which has the potential to address the issues of Fe-FET in non-volatile memory applications. In our FeS-FET, a ferroelectric semiconductor is employed as the channel material while the gate insulator is the dielectric (Fig. 1b). The two non-volatile polarization states in the FeS-FETs exist in the ferroelectric semiconductor. Therefore, a high quality amorphous gate insulator can be used instead of the common polycrystalline ferroelectric insulator found in Fe-FETs. Furthermore, the mobile charges in the semiconductor can screen the depolarization field across the semiconductor. Thus, the charge trapping and leakage current through the ferroelectric insulator found in conventional Fe-FETs can be potentially eliminated. As a result, our approach could offer performance improvements over conventional Fe-FETs in non-volatile memory applications.

Our FeS-FETs use the two-dimensional ferroelectric semiconductor α-$In_2Se_3$ as the channel material. α-$In_2Se_3$ was selected as the channel ferroelectric semiconductor[19–25] because of its appropriate bandgap of ~1.39 eV, room temperature ferroelectricity[23] with a Curie temperature above 200 °C , the ability to maintain ferroelectricity down to a few atomic layers[20,22] and the feasibility of large-area growth[26,27]. When using a scaled $HfO_2$ gate insulator, the fabricated FeS-FETs exhibit high performance with a large memory window, a high on/off ratio of over $10^8$, a maximum on-current of 862 μA/μm, and a low supply voltage.

**FeS-FET Device Physics**

The working mechanism of a FeS-FET is fundamentally different from a traditional Fe-FET. In a Fe-FET, only the polarization bound charges at the gate insulator/semiconductor interface can affect the electrostatics. The ferroelectric polarization switching can tune the threshold voltage ($V_T$) of the device by reversing the polarity of the polarization bound charge, thereby a counterclockwise $I_D$-$V_{GS}$ hysteresis loop can be achieved. However, in a FeS-FET, the polarization charges accumulate at both bottom surface (BS) and top surface (TS) of the ferroelectric semiconductor, as shown in Fig. 1c. As a result, the drain current ($I_D$) of the FeS-FET is determined by both bottom surface and top surface of the semiconductor. As shown in Fig. 1c, if the FeS-FET is in polarization down state, negative bound charges accumulate at the top surface while positive bound charges accumulate at the bottom surface, and vice versa for polarization up state.

The coupled ferroelectric and semiconducting nature of α-In$_2$Se$_3$ is critical to analyze and understand the device operation of the FeS-FET. Due to the semiconducting nature, the channel can exhibit mobile charges depending on the relative positions of conduction/valence band edges and the Fermi level ($E_F$). The presence of such mobile charges can allow a non-uniform distribution of electric field (E field) across the different layers of α-In$_2$Se$_3$. The amount of E field in the channel, in turn, determines the extent of polarization switching in α-In$_2$Se$_3$ layers. Such complex interactions in the proposed device lead to unique characteristics, which can be broadly categorized into clockwise hysteretic and counterclockwise hysteretic. The direction of hysteresis is dependent on E field in the channel, determined by the gate dielectric thickness and the applied gate voltage ($V_{GS}$). To explain this in more details, two different effective oxide thicknesses (EOT) are considered: high EOT and low EOT.

For simplicity, the band diagram of the FeS-FET is firstly discussed without considering band bending induced by semiconducting nature. The major difference between high EOT device and low EOT device is the strength of the electric field across the semiconductor. For example, in the experiments in this work, the maximum voltage applied for high EOT device with 90 nm $SiO_2$ gate insulator is 50 V (EOT~90 nm, max voltage/EOT~0.56 V/nm). The maximum voltage applied for low EOT device with 15 nm $HfO_2$ gate insulator is 5 V (EOT~3 nm, max voltage/EOT~1.7 V/nm). The huge difference on the displacement field (D field) is critical to the I-V characteristics of the FeS-FET. For high EOT devices, electric field across the semiconductor is not strong enough to penetrate to the top surface of the ferroelectric semiconductor, as shown in Fig. 1d. Therefore, only partial switching happens near the oxide/semiconductor interface. For low EOT devices, the electric field is sufficiently strong to trigger full polarization switching in the ferroelectric semiconductor, as shown in Fig. 1e. The high and low EOT conditions are applied for the following discussion. Numerical simulation confirms the validity all these conditions and will be discuss later.

In high EOT condition, as shown in the band diagram in Fig. 1d, in polarization down state, mobile charges can accumulate at the bottom surface because of the band bending so that channel resistance is low. Similarly, in polarization up state, mobile charges density at the bottom surface is low, resulting high channel resistance. A negative voltage below the coercive voltage ($V_{C-}$) leads to polarization down state and a positive voltage above the coercive voltage ($V_{C+}$) leads to polarization up state. Therefore, drain current in high EOT condition ($I_{D,HE}$) versus $V_{GS}$ curve has a clockwise hysteresis loop. In low EOT condition, the electric field across the semiconductor is sufficiently large so that it can penetrate fully into the semiconductor. In this case, top surface can become conducting due to full polarization switching. As shown in the band diagram in Fig. 1e, in polarization down state, mobile charges do not exist at the top surface so that the channel resistance is high. Similarly, in polarization up state, mobile charges can accumulate at the top surface, resulting low channel resistance. Note that the bottom surface is much easier to be controlled by the gate voltage so that by adjusting the gate voltage the mobile charge at bottom surface is fully depleted. Therefore, drain current in low EOT condition ($I_{D,LE}$) versus $V_{GS}$ curve has a counterclockwise hysteresis loop. Therefore, both directions of hysteresis loop can be realized depending on EOT of the dielectric. Note that the above discussion is a simplified picture without considering the band bending induced by mobile carriers. A sophisticated and complete theory and numerical simulation considering both ferroelectric and semiconducting nature are developed showing the same conclusion and will be discussed later. Experimentally, α-In$_2$Se$_3$ FeS-FETs with 90 nm SiO$_2$ as gate insulator as in a high EOT condition show a clockwise hysteresis loop while α-In$_2$Se$_3$ FeS-FETs with 15 nm HfO$_2$ as in a low EOT condition show a counterclockwise hysteresis loop.

**Experiments on α-In$_2$Se$_3$ FeS-FETs**

α-In$_2$Se$_3$ is a recently discovered 2D ferroelectric semiconductor, which is employed in this work to demonstrate the FeS-FET operation. α-In$_2$Se$_3$ bulk crystals were grown by melt method with a layered non-centrosymmetric rhombohedral R3m structure[19], as shown in Fig. 2a. The α-In$_2$Se$_3$ FeS-FET (as shown in Fig. 2b) consists a heavily p-doped Si substrate as back-gate electrode, 15 nm HfO$_2$ or 90 nm SiO$_2$ as the gate insulator, 2D α-In$_2$Se$_3$ as ferroelectric semiconductor channel and 30 nm Ti/50 nm Au as source/drain electrodes. An optimized 10 nm 175 °C-grown ALD Al$_2$O$_3$ capping layer is grown on top of the α-In$_2$Se$_3$ channel, which gives a significantly performance enhancement comparing with α-In$_2$Se$_3$ FeS-FETs without passivation. High-angle annular dark field STEM (HAADF-STEM) image of thin α-In$_2$Se$_3$ flake is shown in Fig. 2c. Distinct arrangement of atoms could be clearly identified, with the fringe space of (100) planes measured to be 0.35 nm, confirming an ideal hexagonal lattice structure of α-In$_2$Se$_3$. The corresponding selected area electron diffraction (SAED) showing 6-fold symmetry with perfect hexagonal crystal structure, indicating the α-In$_2$Se$_3$ flake is highly single-crystallized. The corresponding spectrum of Energy Dispersive Spectroscopy (EDS) is shown in Fig. 2d, which confirms the atomic percentage (at %) ratio between In and Se is ~ 2:3. Fig. 2e shows a photoluminescence (PL) spectrum of a bulk α-In$_2$Se$_3$ crystal, measured from 1.2 eV to 1.6 eV, indicating a direct bandgap of ~1.39 eV. Fig. 2f shows a Raman spectrum measured from a bulk α-In$_2$Se$_3$ crystal, showing consistent peak positions comparing to literature reports[24].

A strong piezoelectric response is observed from a 78.7 nm thick α-In$_2$Se$_3$ flake, which is shown in Fig. 3a-c by piezoresponse force microscopy (PFM). To extract the piezoelectric coefficient, different AC voltages are applied on the sample from the conductive atomic force microscopy (AFM) tip which shows a linear relationship between the mechanical deformation (PFM amplitude) and the electric field. The effective piezoelectric coefficient ($d_{33}$) of the α-In$_2$Se$_3$ flake is 32 pm/V (Supplementary section 1). From the background noise shown in Fig. 3c, it is clear that the mechanical deformation in the PFM measurement is dominated by the intrinsic ferroelectric polarization. Fig. 3d and 3g show two different measurement schematics on α-In$_2$Se$_3$, the metal-semiconductor-metal (MSM) structure and the metal-oxide-semiconductor (MOS) structure. Fig. 3e and 3f show PFM phase and PFM amplitude versus voltage hysteresis loop of a 15.3 nm thick α-In$_2$Se$_3$ flake on conductive Ni, showing clear ferroelectric polarization switching under external electric field. The PL measurement of bandgap and PFM measurement of polarization switching together suggest the α-In$_2$Se$_3$ used in this work is a ferroelectric semiconductor. Note that PFM hysteresis is just a necessary condition but is not a sufficient condition for ferroelectric materials. As mobile charges exist in a semiconductor, such charges may screen and prevent the electric field to penetrate into the body of the semiconductor, so that the ferroelectric polarization switching may be different in a MOS structure. Therefore, it is important to test whether the polarization in α-$In_2Se_3$ can be switched by an external electric field in a MOS structure. Fig. 3h and 3i show PFM phase and PFM amplitude versus voltage hysteresis loop of 6 nm $Al_2O_3$/16.3 nm α-$In_2Se_3$ on Ni. The ferroelectric hysteresis loop suggests that α-$In_2Se_3$ has switchable polarization in a MOS device structure. Thus, it is viable to apply α-$In_2Se_3$ as the channel for a FeS-FET. The raw data of PFM measurement can be found in supplementary section 1.

Fig. 4a illustrates a top-view false-color scanning electron microscope (SEM) image of a fabricated α-$In_2Se_3$ FeS-FET with ALD passivation, capturing the α-$In_2Se_3$ thin film and the Ti/Au electrodes. The electrical performance of unpassivated device can be found in supplementary section 2. The α-$In_2Se_3$ FeS-FET without ALD passivation shows clear clockwise hysteresis loop and a large memory window over 70 V. A high on/off ratio over $10^7$ at $V_{DS}$=0.5 V between on- and off-states is also achieved. It is found the performance of the α-$In_2Se_3$ FeS-FET can be further enhanced by ALD $Al_2O_3$ passivation, as shown in Fig. 4a. Fig. 4b shows the $I_D$-$V_{GS}$ characteristics of a representative α-$In_2Se_3$ FeS-FET with ALD passivation and 90 nm $SiO_2$ as gate insulator. Low-temperature grown ALD $Al_2O_3$ not only offers passivation on α-$In_2Se_3$ surface, but also provides electron doping effect due to the positive fixed charges.[28,29] The transfer curve is measured by double gate voltage sweep and at different $V_{DS}$. The device has a channel length ($L_{ch}$) of 1 μm, channel thickness ($T_{ch}$) of 52.2 nm. The transfer curve shows clear clockwise hysteresis loop. A high on/off ratio over $10^8$ at $V_{DS}$=1 V is also achieved, suggesting a high quality oxide/semiconductor interface. The minimum subthreshold slope (SS) at $V_{DS}$=0.05 V achieved in this device is 650 mV/dec, indicating an estimated interface trap density ($D_{it}$) of 2.6×$10^{12}$ $cm^{-2}$ without considering the semiconductor capacitance. Fig. 4c shows the $I_D$-$V_{DS}$ characteristics of the same α-In$_2$Se$_3$ FeS-FET as in Fig. 4b. A maximum drain current of 671 µA/µm is achieved. Considering the long channel length (L$_{ch}$=1 µm) used here, the α-In$_2$Se$_3$ FeS-FETs can have much higher on-current at shorter channel length and has the potential for high speed applications. Fig. 4d shows the g$_m$-V$_{GS}$ characteristics at V$_{DS}$=0.05 V of the same device as in Fig. 4b. Maximum g$_m$ at V$_{DS}$=0.05 V of 0.60 µS/µm and 0.94 µS/µm are obtained for forward and reverse gate voltage sweeps, respectively. The extrinsic field-effect mobility (µ$_{FE}$) is calculated using maximum g$_m$ in forward sweep to be 312 cm$^2$/V·s and in reverse sweep to be 488 cm$^2$/V·s without extracting the relative large contact resistance due to the Schottky contacts. Note that the extrinsic field-effect mobility is different from the intrinsic mobility of α-In$_2$Se$_3$. Here, it only serves as a reference for transport properties of the devices (A discussion on the accuracy of field-effect mobility estimation is in supplementary section 5). The performance of the α-In$_2$Se$_3$ FeS-FETs are significantly improved by the 10 nm Al$_2$O$_3$ ALD passivation, comparing to the unpassivated devices shown in supplementary section 2 (µ$_{FE}$=19.3 cm$^2$/V·s in forward sweep and µ$_{FE}$=68.1 cm$^2$/V·s in reverse sweep). The α-In$_2$Se$_3$ FeS-FET with 90 nm SiO$_2$ gate insulator is also characterized at very low temperature down to 80 mK. The existence of hysteresis window similar to room temperature one indicates that the clockwise I$_D$-V$_{GS}$ hysteresis loop is caused by ferroelectric polarization switching instead of charge trapping[30,31]. A detailed discussion is in supplementary section 3. Note that the clockwise hysteresis properties in the FeS-FETs can be integrated together with Fe-FETs as a Fe$^2$-FET, where both insulator and semiconductor are ferroelectric. In the Fe$^2$-FET, a deep steep-slope subthreshold and hysteresis-free can be achieved at the same time (see supplementary section 8 for details) if all the device parameters are optimized.

The clockwise hysteresis loop in α-In$_2$Se$_3$ FeS-FETs indicates the bottom surface is the dominating conducting channel (Fig. 1d). The counterclockwise hysteresis loop can be achieved by careful design of the device structure. By scaling the gate oxide thickness and applying high-k dielectrics (HfO$_2$), a much higher D field can be applied inside the gate oxide. By applying high D field, the I$_D$-V$_{GS}$ curve of α-In$_2$Se$_3$ FeS-FET with 15 nm HfO$_2$ as gate dielectric becomes counterclockwise (and also with high on/off ratio > 10$^8$), as shown in Fig. 4e. The device has a channel length of 1 μm and a channel thickness of 79 nm. The enhanced electric field inside α-In$_2$Se$_3$ can penetrate through to the top surface so that full polarization switching happens instead of partial switching in 90 nm SiO$_2$ case. As a result, top surface conduction can lead to a counterclockwise hysteresis loop, as shown in Fig. 1e. As charge trapping process cannot lead to a counterclockwise hysteresis loop, this result serves as a conclusive proof of the existence of ferroelectricity and polarization switching. Fig. 4f shows the I$_D$-V$_{DS}$ characteristics of an α-In$_2$Se$_3$ FeS-FET. The device has a channel length of 1 μm, channel thickness of 92.1 nm. A maximum drain current of 862 μA/μm is achieved. The devices with 15 nm HfO$_2$ as gate dielectrics and ALD passivation exhibit a significantly reduced supply voltage compared to the devices with 90 nm SiO$_2$ as gate insulator, suggesting the potential of α-In$_2$Se$_3$ FeS-FETs for low power non-volatile memory application.

**Simulation of α-In$_2$Se$_3$ FeS-FETs**

Theory of FeS-FETs and device level simulations have been conducted to investigate the clockwise and counterclockwise hysteresis in the I-V characteristics of FeS-FET. More specifically, physics based self-consistent simulation of FeS-FET devices is performed by coupling Poisson's equation, Ginzburg-Landau equation and 2D charge equation. A van der Waals gap is assumed between the source/drain contacts and the semiconductor channel because of the 2D layered nature of α-In$_2$Se$_3$. The detailed simulation methods can be found in supplementary section 6. As shown in the simulation results in Fig. 5a and 5c, a clockwise $I_D$-$V_{GS}$ hysteresis loop is achieved for high EOT (30 nm) device and a counterclockwise hysteresis loop is achieved for low EOT (0.5 nm) device. For low EOT device, at $V_{GS}$=-1 V in forward gate voltage sweep, the device is in polarization down state. As shown in the band diagram in Fig. 5b, mobile charges density in the semiconductor is low so that the device is in off-state. At $V_{GS}$=-3.24 V in reverse gate voltage sweep, the device is in polarization up state. This is a full polarization switching as shown in the polarization vector map. For high EOT device, at $V_{GS}$=-20 V in forward gate voltage sweep, the device is in polarization down state. The device is in on-state due to bottom surface inversion. At $V_{GS}$=-10 V in reverse gate voltage sweep, the ferroelectric semiconductor is in polarization up state only near the gate oxide/semiconductor interface, indicating this is a partial polarization switching. As a result, the device is in off-state without top surface conduction, as shown in Fig. 5d. A detailed analysis by band diagrams at different gate voltages during bi-directional gate voltage sweep for both low EOT and high EOT is discussed in supplementary section 6. The simulation results considering both ferroelectric and semiconducting nature of the α-In$_2$Se$_3$ confirm the validity of the simple picture discussed in Fig. 1.

**Conclusion**

We have reported a FeS-FET in which the 2D ferroelectric semiconductor α-In$_2$Se$_3$ is used as channel material. An ALD Al$_2$O$_3$ passivation method was developed to protect and enhance the performance of the α-In$_2$Se$_3$ FeS-FETs. The fabricated FeS-FETs exhibit high performance with a large memory window, a high on/off ratio over $10^8$, a maximum on-current of 862 µA/µm, and low supply voltage. Our FeS-FETs have the potential to surpass the capabilities of existing Fe-FETs for non-volatile memory applications.

## Methods

**Device Fabrication.** α-In$_2$Se$_3$ were transferred onto a 15 nm HfO$_2$ or 90 nm SiO$_2$ on Si substrate using the Scotch tape exfoliation. The p+ Si wafers with 90 nm thermal grown SiO$_2$ were purchased from WaferPro, LLC. 15 nm HfO$_2$ was deposited by ALD using [(CH$_3$)$_2$N]$_4$Hf (TDMAHf) and H$_2$O as precursors at 200 °C. 30 nm Ti and 50 nm Au were deposited by electron-beam evaporation and followed by a lift-off process for α-In$_2$Se$_3$ back-gate transistors. An optimized 10 nm Al$_2$O$_3$ was finally deposited by ALD using Al(CH$_3$)$_3$ (TMA) and H$_2$O as precursors at 175 °C.

**Material Characterization.** Material characterizations on α-In$_2$Se$_3$ crystals were carried out to investigate α-In$_2$Se$_3$ as a single crystal, semiconducting and ferroelectric material, including STEM, photoluminescence, Raman spectroscopy and piezoresponse force microscopy. HAADF-STEM were performed with FEI Talos F200x equipped with a probe corrector. This microscope was operated with an acceleration voltage of 200 kV. Raman and photoluminescence measurements were carried out on a HORIBA LabRAM HR800 Raman spectrometer. DART-PFM was carried out on Asylum Cypher ES. Single-phase PFM characterization was carried out on Keysight 5500 under the contact mode and the conductive AFM tip has averaged spring constant ~5N/m.

**Device Characterization.** The thickness of the α-In$_2$Se$_3$ was measured using a Veeco Dimension 3100 AFM system. SEM and EDS analysis were done using a Hitachi S-4800 FE-SEM and an Oxford X-Max Silicon Drift Detector. DC electrical characterization was performed with a Keysight B1500 system in dark environment. Electrical data was collected with a Cascade Summit probe station at room temperature. Low temperature Hall measurement at 80 mK was performed in an Oxford Triton 300 dilution fridge.

**Data availability.** The data that support the plots within this paper and other findings of this study are available from the corresponding author upon reasonable request.

**Acknowledgements**

The work was supported in part by NSF/AFOSR EFRI 2DARE program and in part by ASCENT, one of six centers in JUMP, a Semiconductor Research Corporation (SRC) program sponsored by DARPA. J.J. and H.W. acknowledge the support from the U.S. Office of Naval Research for the TEM effort at Purdue.

**Supplementary Information**

Additional details for effective piezoelectric constant and DART-PFM raw data, device performance for unpassivated FeS-FETs, low temeprature Hall measurement at 80 mK, channel thickness dependence, discussions on the acuracy of field-effect mobility, theory and simulation of FeS-FETs, the discussion of partial switching in low EOT FeS-FETs, and a deep steep-slope and hysteresis-free transistor concept based on a ferroelectric semiconductor are in the supplementary information.

**Author Contributions**

P.D.Y. and M.S. conceived the idea and proposed the FeS-FET concept. M.S. did the device fabrication, electrical measurement and analysis. A.K.S and S.K.G. did the numerical simulation.

S.G. and W.W. performed the PFM measurements. J.Q., J.J and H.W. conducted the TEM and EDS measurement. Y.D. and M.S. did the SEM imaging and EDS analysis. G.Q. did the Raman and PL measurements. G.Q. and C.N. performed the low temperature I-V and Hall measurements. M.S. and P.D.Y. co-wrote the manuscript and all authors commented on it.

**Financial Interest Statement**

The authors declare no competing financial interest.

**Figure captions**

**Figure 1 | Schematic diagram and proposal of a ferroelectric semiconductor field-effect transistor. a,** Schematic diagram of a ferroelectric field-effect transistor (Fe-FET). **b,** Schematic diagram of a ferroelectric semiconductor field-effect transistor (FeS-FET). In the FeS-FET, the conventional semiconductor channel is replaced by a ferroelectric semiconductor, while the gate insulator is still conventional dielectric. **c,** Polarization bound charge distribution in FeS-FET in polarization down (after negative gate bias) and polarization up (after positive gate bias) states. **d,** Band diagram of FeS-FET with high EOT in polarization up and polarization down states and the corresponding $I_D$-$V_{GS}$ characteristics. A clockwise hysteresis loop is achieved due to partial polarization switching. **e,** Band diagram of FeS-FET with low EOT in polarization up and polarization down states and the corresponding $I_D$-$V_{GS}$ characteristics. A counterclockwise hysteresis loop is achieved due to full polarization switching.

**Figure 2 | Material properties of ferroelectric semiconductor α-In$_2$Se$_3$. a,** Crystal structure of ferroelectric semiconductor α-In$_2$Se$_3$. **b,** Schematic diagram of the experimental α-In$_2$Se$_3$ FeS-FET. The experimental α-In$_2$Se$_3$ FeS-FET consists heavily-doped silicon substrate as back-gate electrode, 15 nm HfO$_2$ or 90 nm SiO$_2$ as gate dielectric, 2D thin-film α-In$_2$Se$_3$ as the channel

ferroelectric semiconductor, 30 nm Ti/50 nm Au as source/drain electrodes. **c,** HAADF-STEM image and the corresponding SAED of thin α-In$_2$Se$_3$ film. **d,** EDS spectrum of thin α-In$_2$Se$_3$ film. The measured atomic percent (at %) of In and Se is 37 and 63, respectively. **e,** Photoluminescence spectrum and **f,** Raman spectrum of bulk α-In$_2$Se$_3$, showing a bandgap of ~1.39 eV without considering 2D exciton binding energy. Raman and photoluminescence spectrums are measured at room temperature and confirm the semiconducting properties of the α-In$_2$Se$_3$.

**Figure 3 | PFM measurement on α-In$_2$Se$_3$ thin film**. **a,** PFM amplitude, **b,** PFM phase and **c,** PFM background images of a 78.7 nm thick α-In$_2$Se$_3$ flake on a heavily doped silicon substrate. **d,** Schematic diagram of PFM measurement using metal-semiconductor-metal (MSM) structure. **e,** PFM phase and **f,** PFM amplitude versus voltage hysteresis loop of 15.3 nm α-In$_2$Se$_3$ on Ni, showing clear ferroelectric polarization switching under external electric field. **g,** Schematic diagram of PFM measurement using metal-oxide-semiconductor (MOS) structure. **h,** PFM phase and **i,** PFM amplitude versus voltage hysteresis loop of 6 nm Al$_2$O$_3$/16.3 nm α-In$_2$Se$_3$ on Ni. The ferroelectric hysteresis loop suggests that α-In$_2$Se$_3$ has switchable polarization in a MOS device structure.

**Figure 4 | Switching characteristics of α-In$_2$Se$_3$ FeS-FETs. a,** Schematic diagram of the experimental α-In$_2$Se$_3$ FeS-FET with ALD passivation and a false-color top-view SEM image of a fabricated α-In$_2$Se$_3$ FeS-FET. **b,** I$_D$-V$_{GS}$, **c,** I$_D$-V$_{DS}$ and **d,** g$_m$-V$_{GS}$ characteristics at room temperature of a representative α-In$_2$Se$_3$ FeS-FET with 90 nm SiO$_2$ as gate dielectrics and ALD passivation. The device has a channel length of 1 μm and channel thickness of 52.2 nm. The device exhibits a large memory window, maximum drain current of 671 μA/μm, on/off ratio > 10$^8$, high electron mobility with μ$_{FE}$=312 cm$^2$/V·s measured in forward sweep and μ$_{FE}$=488 cm$^2$/V·s measured in reverse sweep. **e,** I$_D$-V$_{GS}$ characteristics at room temperature of a representative α-

In$_2$Se$_3$ FeS-FET with 15 nm HfO$_2$ as gate insulator and ALD passivation. The device has a channel length of 1 μm and channel thickness of 79 nm. **f,** I$_D$-V$_{DS}$ characteristics of an α-In$_2$Se$_3$ FeS-FET device with 15 nm HfO$_2$ as gate dielectrics and ALD passivation. The device has a channel length of 1 μm and a channel thickness of 92.1 nm. The device with 15 nm HfO$_2$ as gate dielectrics and ALD passivation exhibits a significantly reduced supply voltage, high on/off ratio $> 10^8$, maximum drain current of 862 μA/μm.

**Figure 5 | Simulation of α-In$_2$Se$_3$ FeS-FETs. a,** Simulation of I$_D$-V$_{GS}$ characteristics of α-In$_2$Se$_3$ FeS-FET with EOT=0.5 nm. **b,** Band diagram at V$_{GS}$=-1 V in forward gate voltage sweep, where the device is in off-state and band diagram at V$_{GS}$=-3.24 V in reverse gate voltage sweep, where the device is in on-state. The counterclockwise hysteresis loop originates from the top surface conduction because of the full polarization switching. **c,** Simulation of I$_D$-V$_{GS}$ characteristics of α-In$_2$Se$_3$ FeS-FET with EOT=30 nm. **d,** Band diagram at V$_{GS}$=-20 V in forward gate voltage sweep, where the device is in on-state and band diagram at V$_{GS}$=-10 V in reverse gate voltage sweep, where the device is in off-state. The polarization vector near the gate insulator/semiconductor interface is zoomed for a better illustration. Note the polarization direction near the interface is opposite to the ones in bulk and surface. The clockwise hysteresis loop originates from the bottom surface conduction because of the partial polarization switching. A detailed analysis by band diagrams at different gate voltages during bi-directional gate voltage sweep for both low EOT and high EOT is discussed in supplementary section 6.

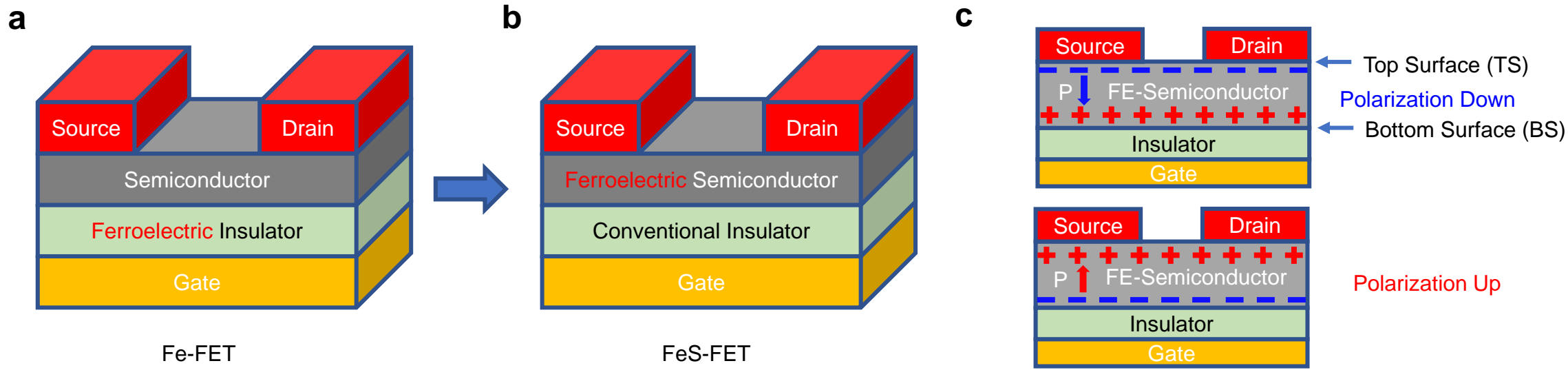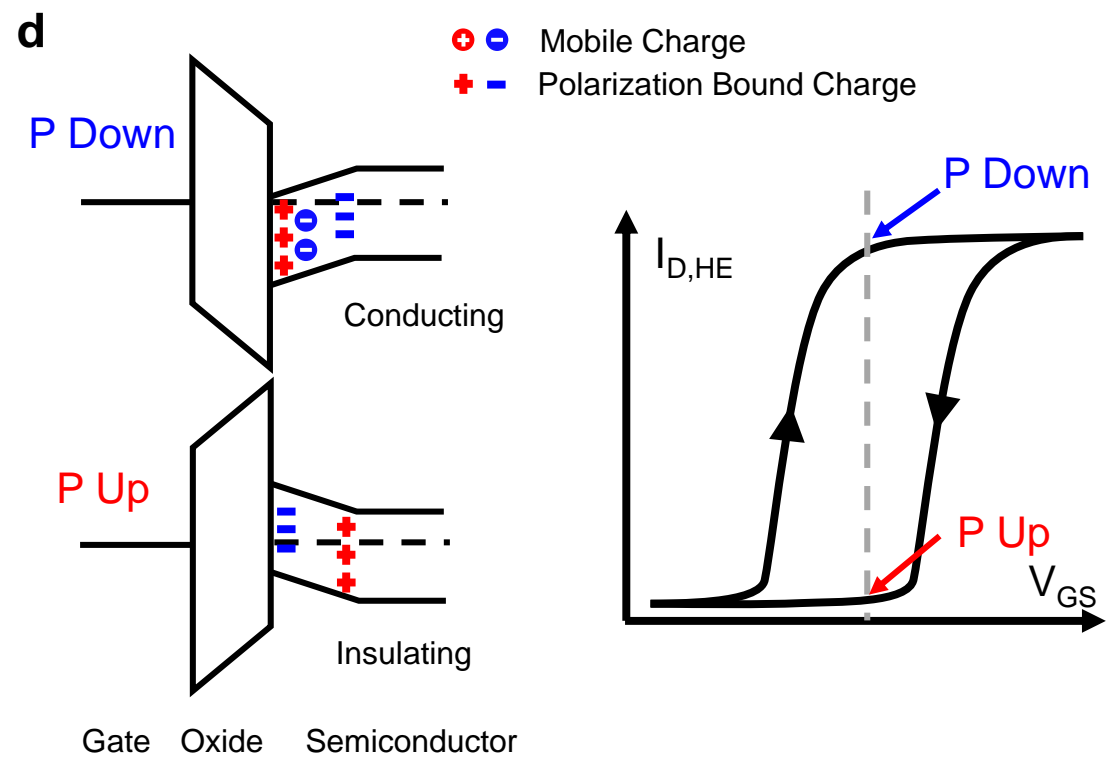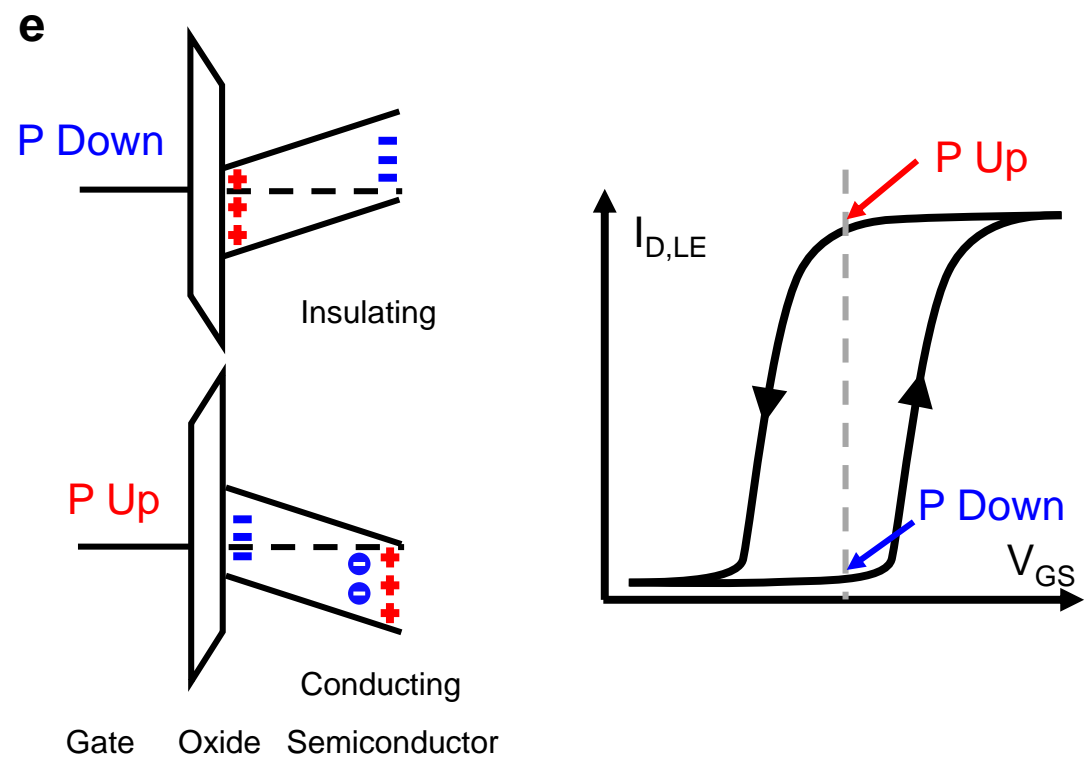

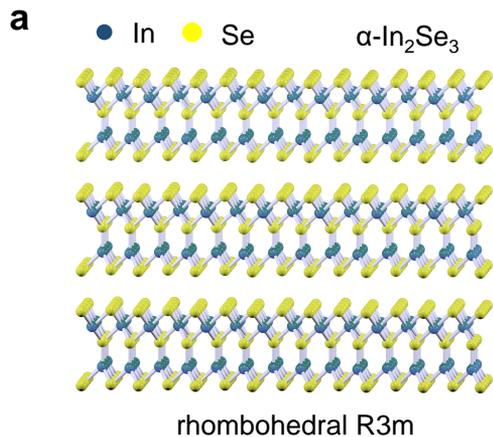
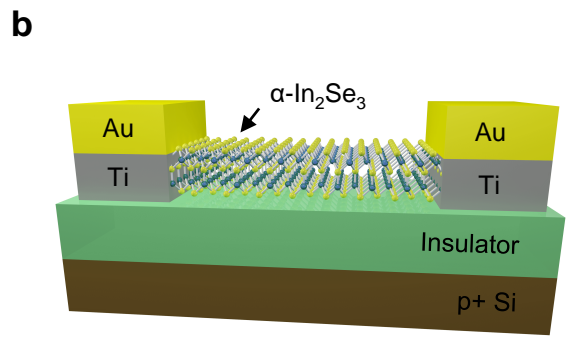
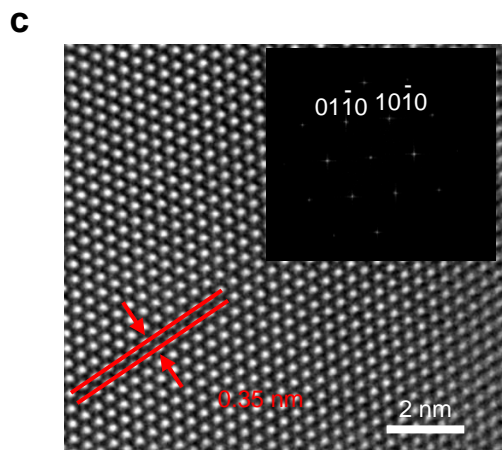
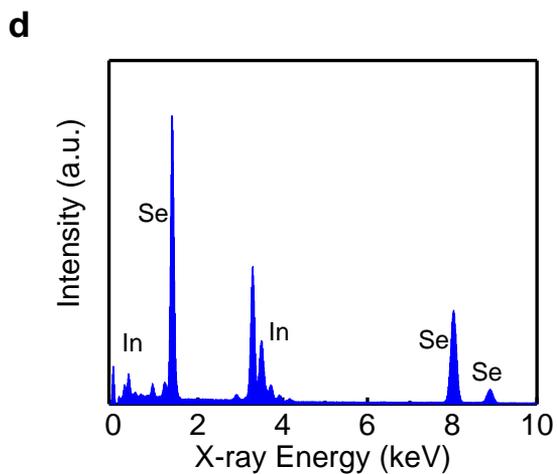
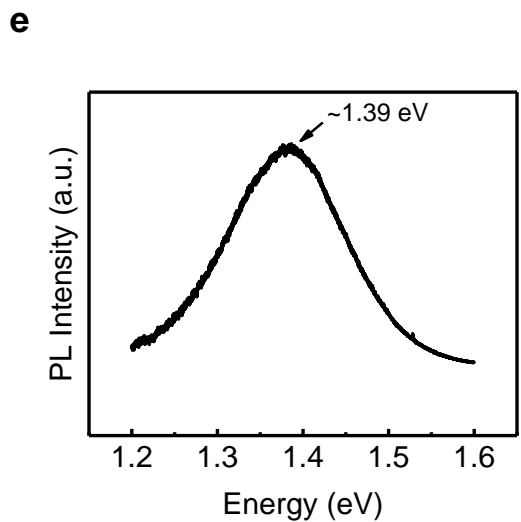
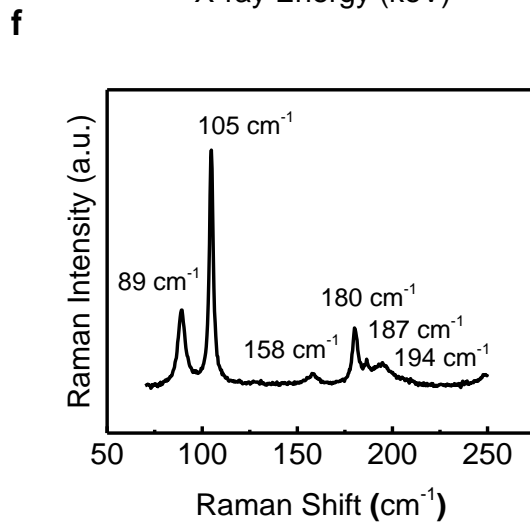

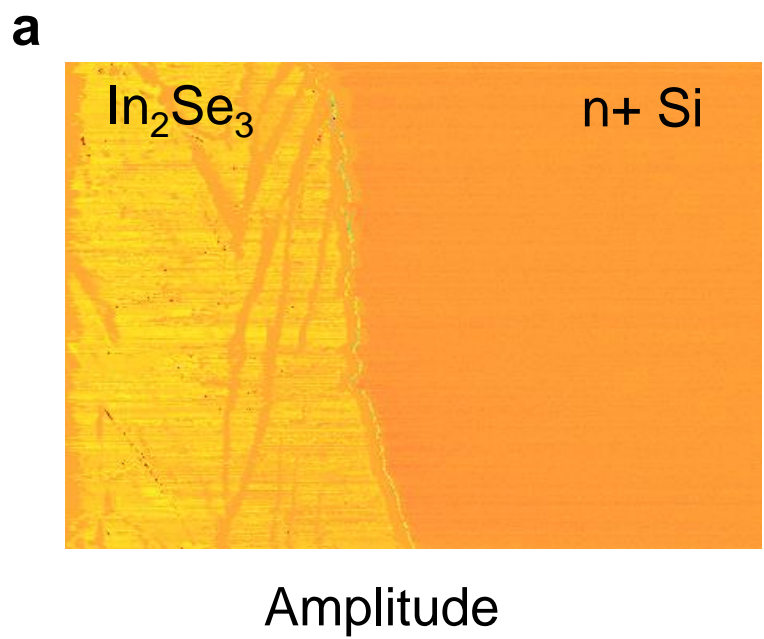 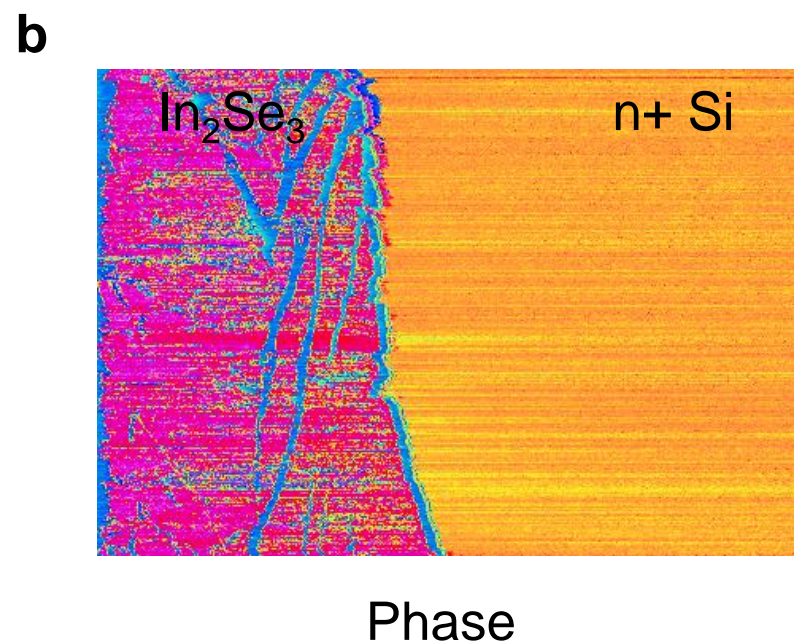 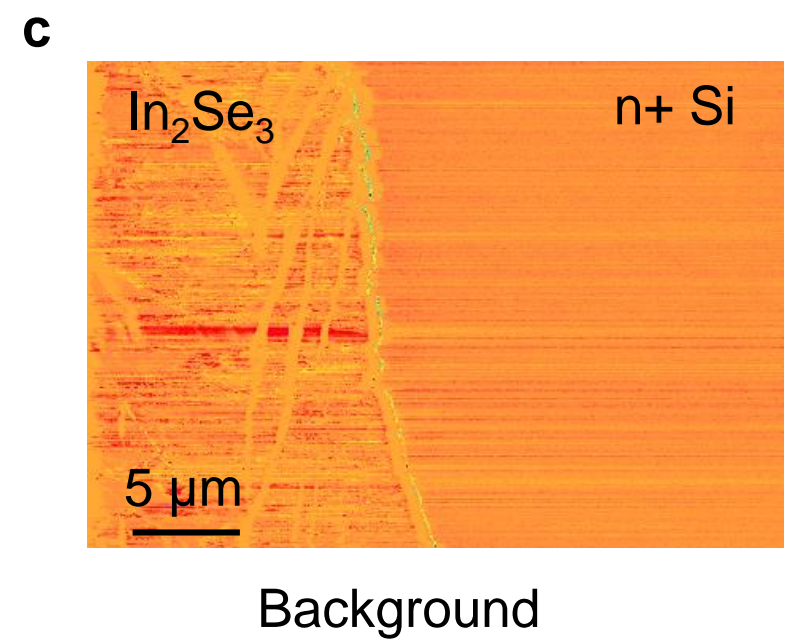

Amplitude | Phase | Background

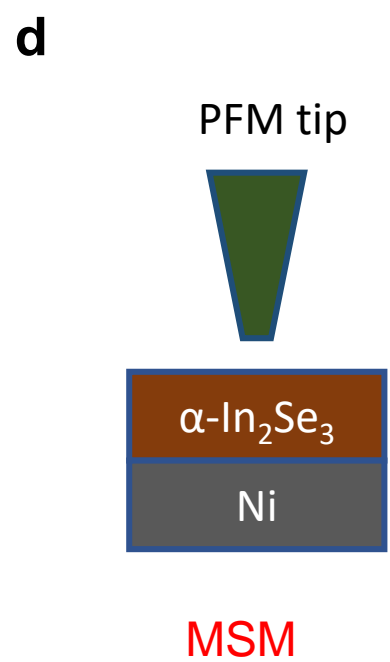 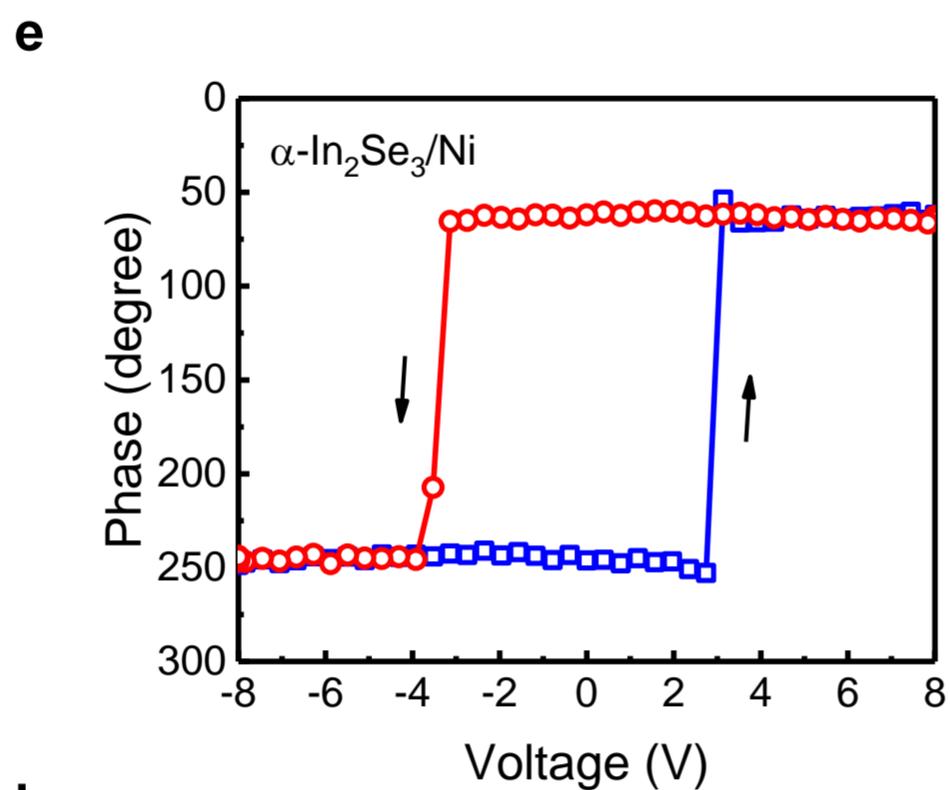 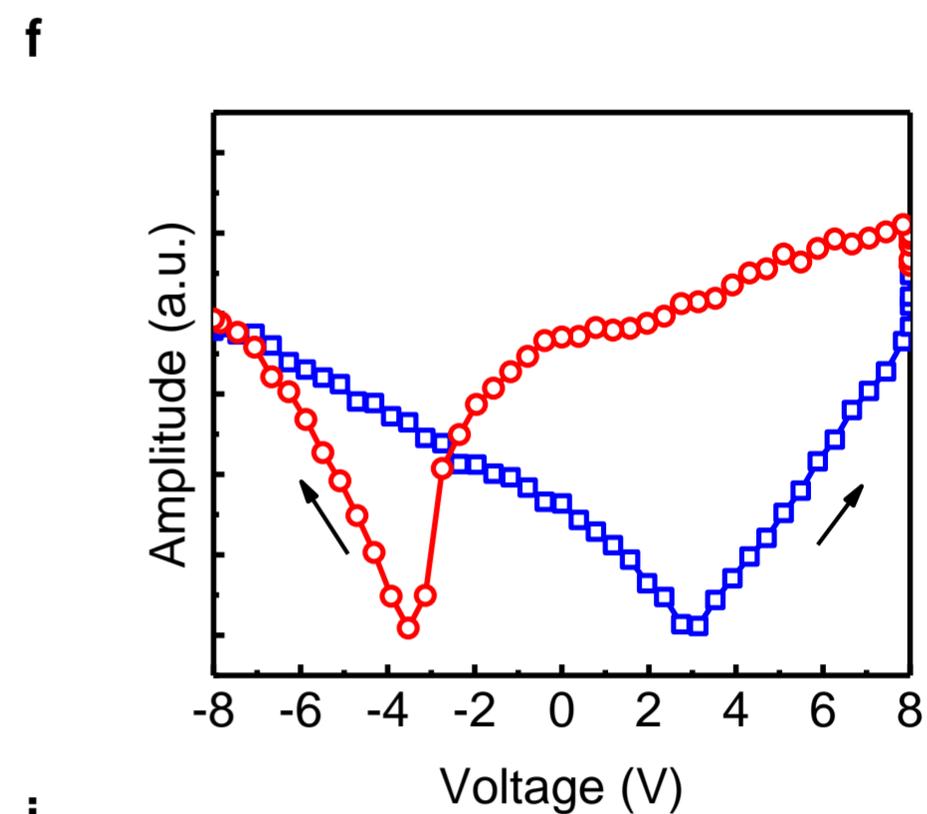

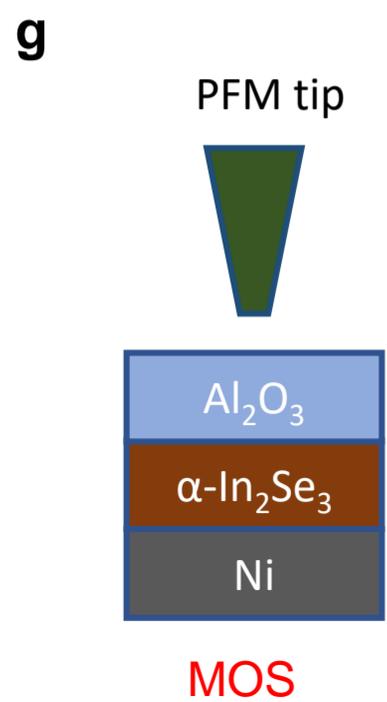 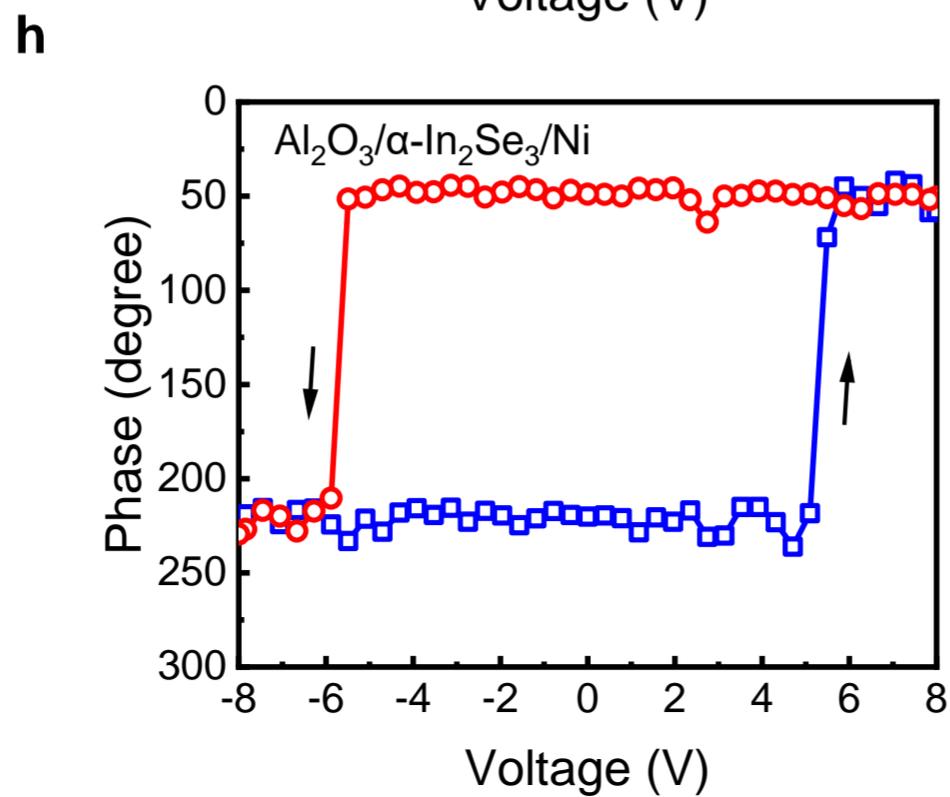 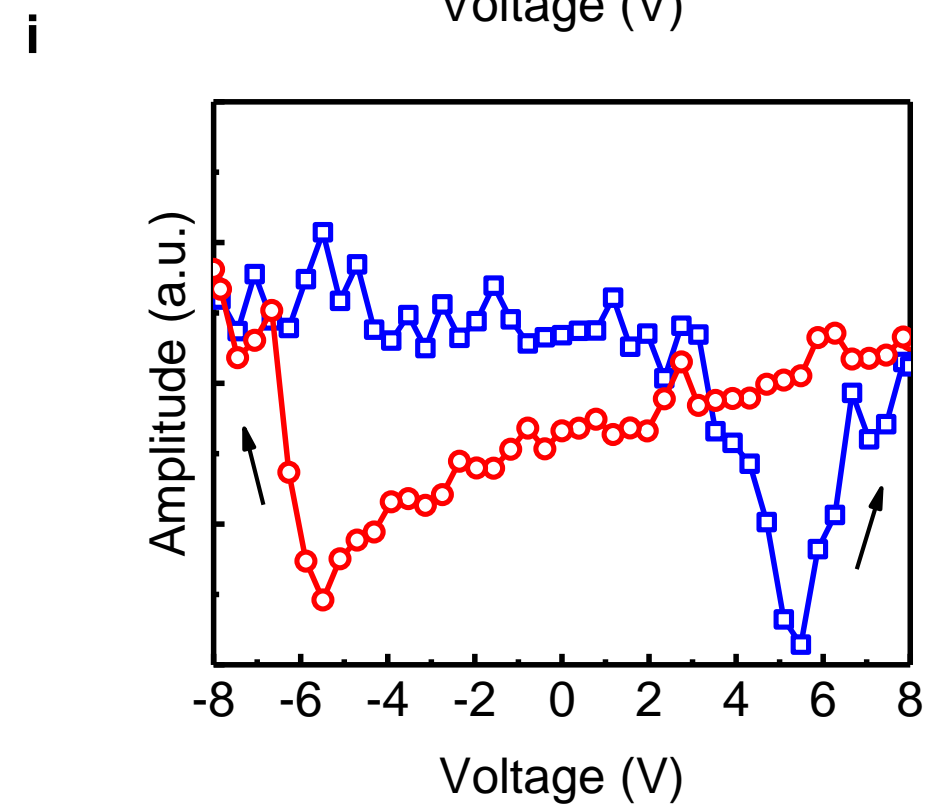

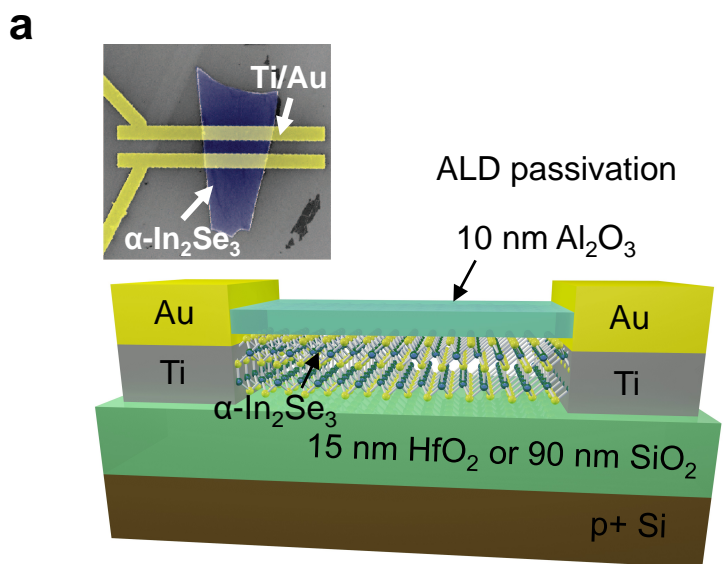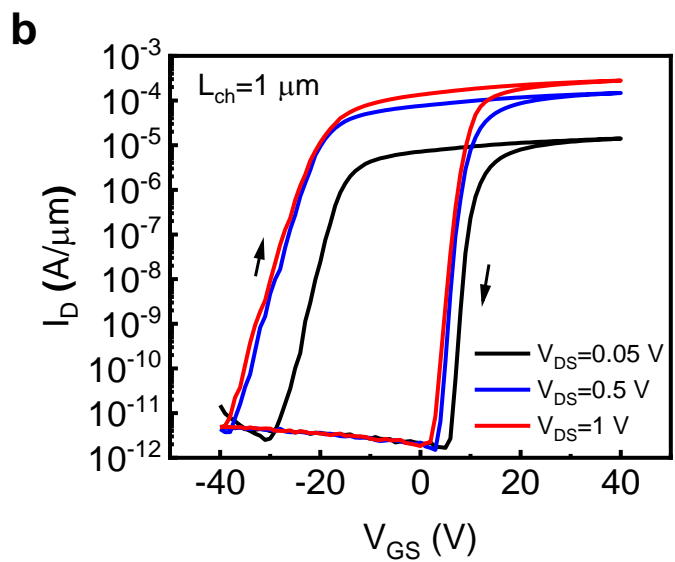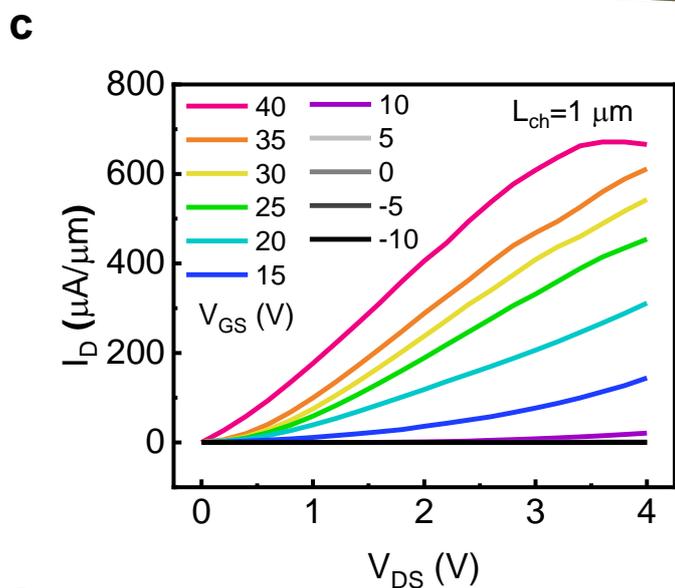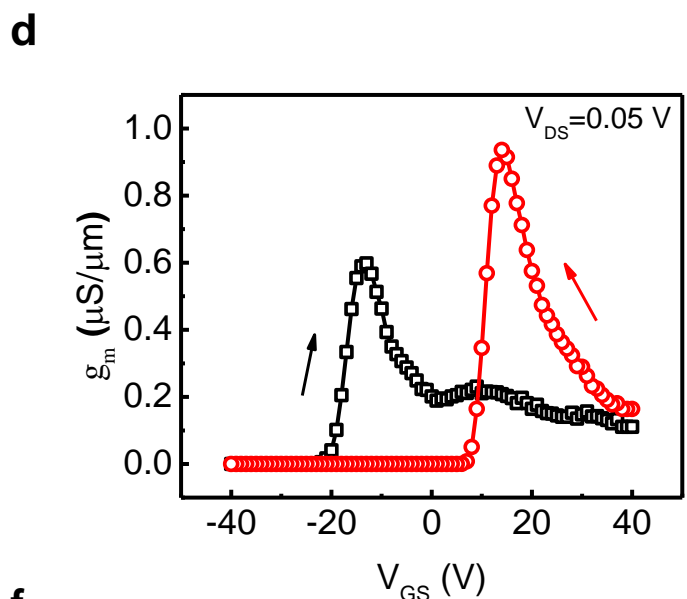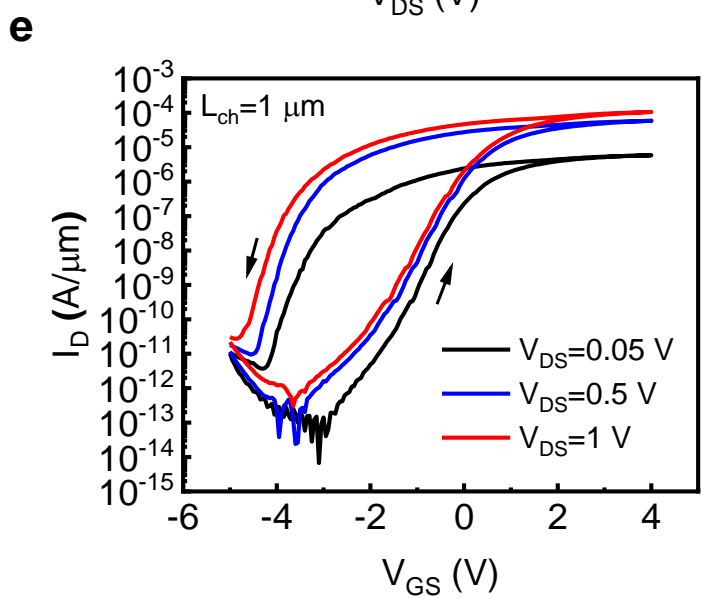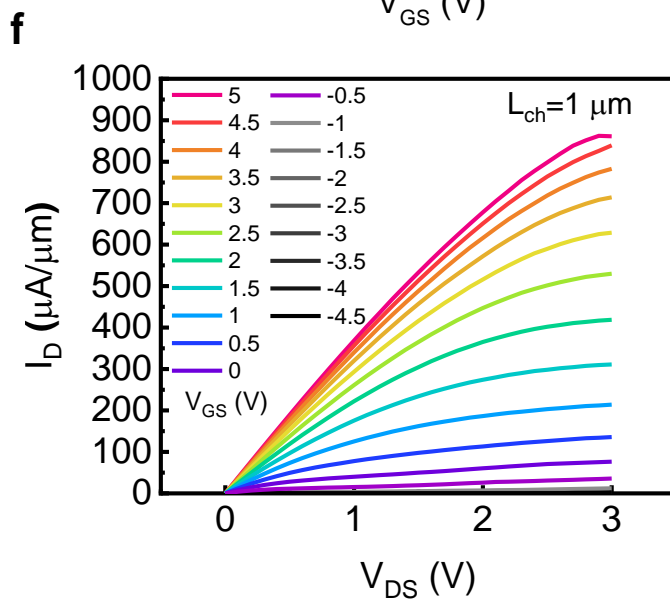

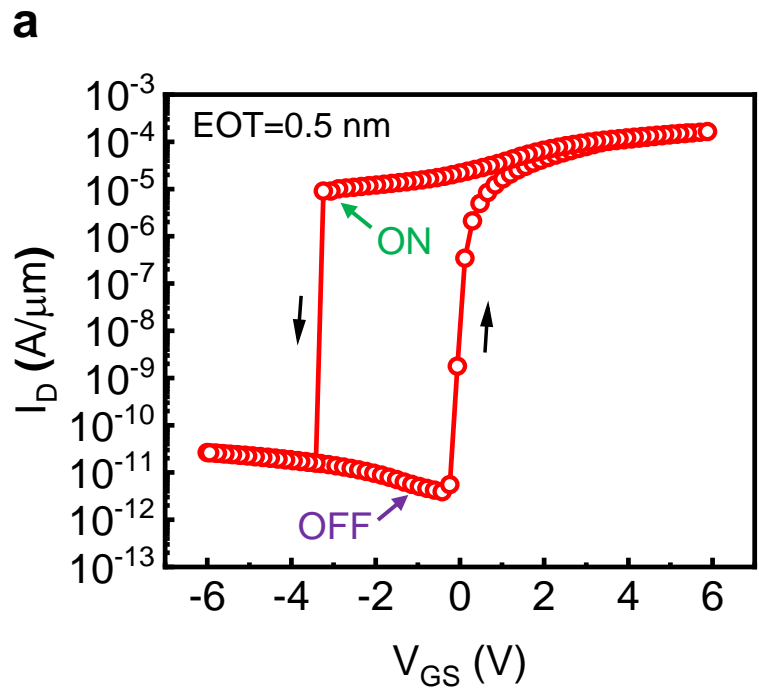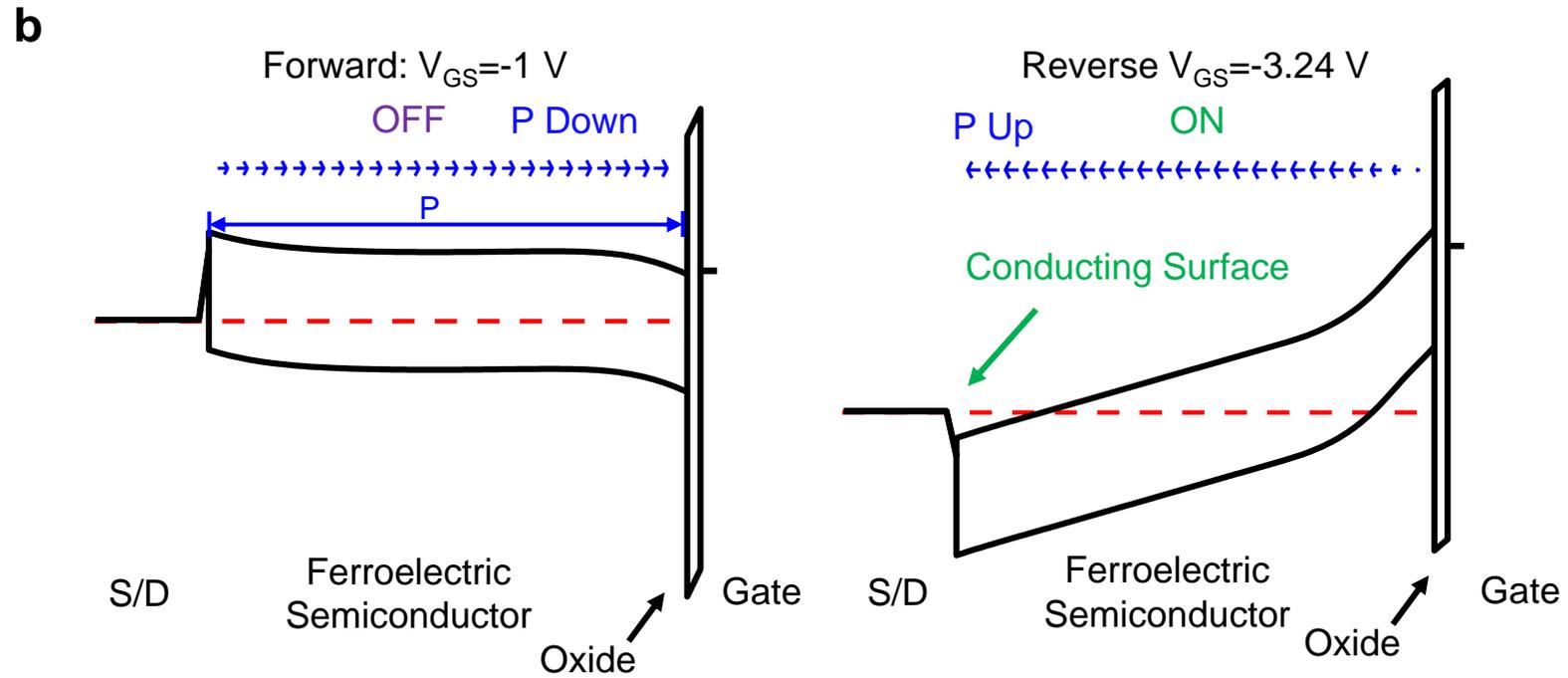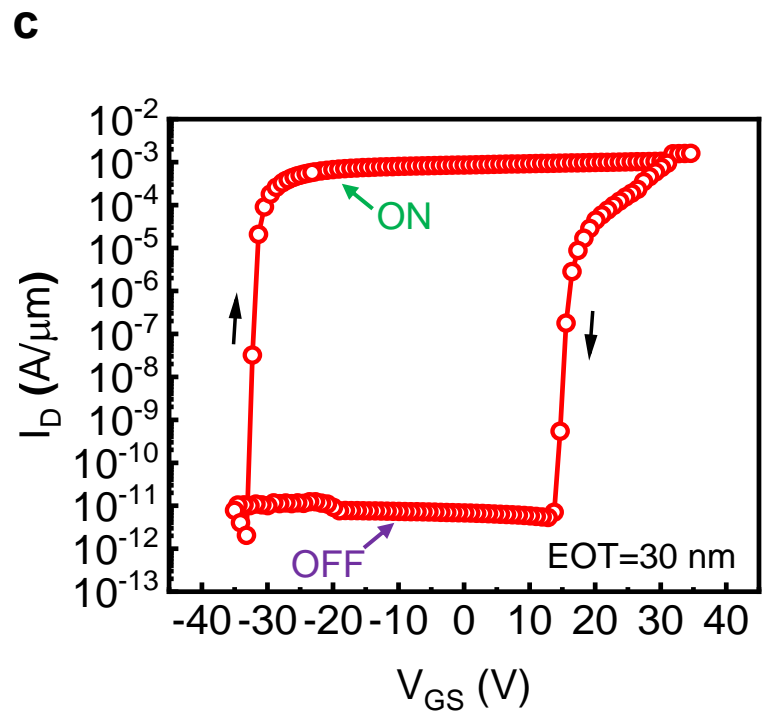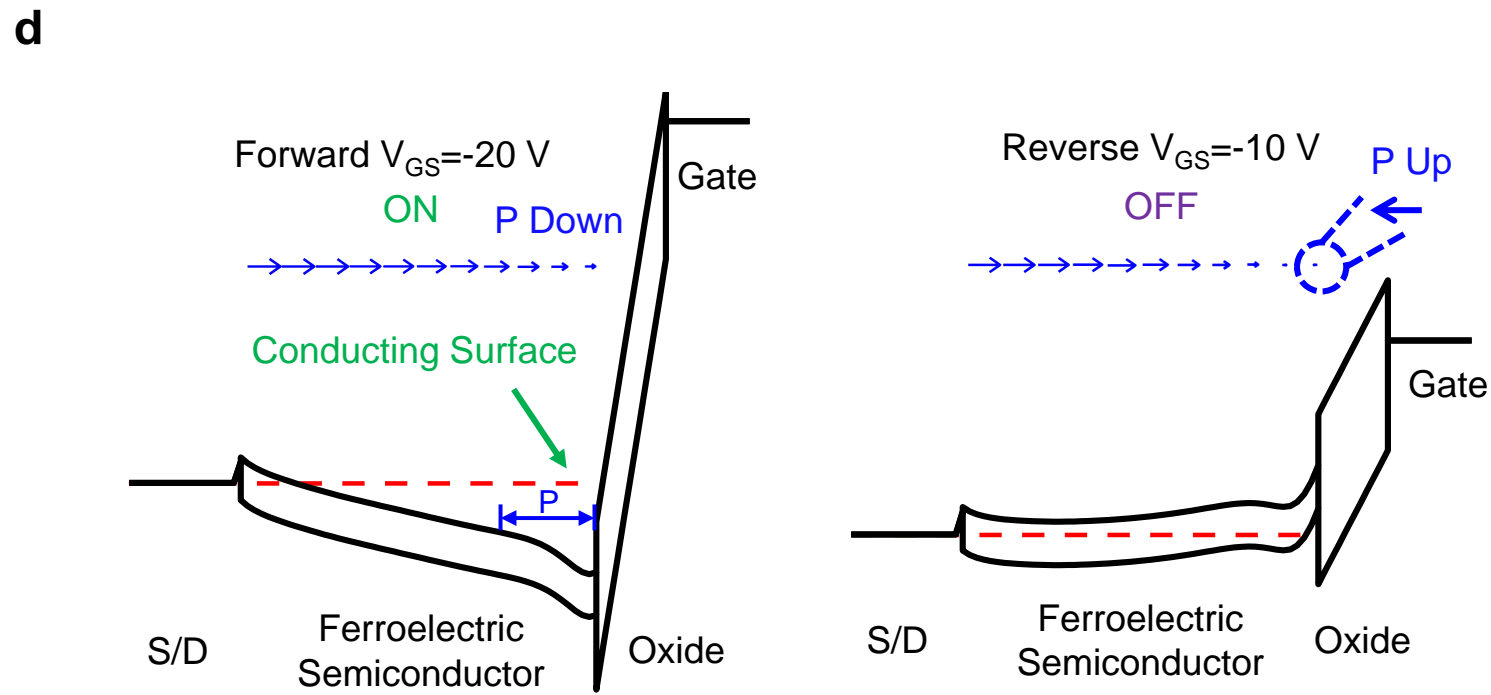

# Supplementary Information

## 1. Piezoelectric coefficient and DART-PFM data

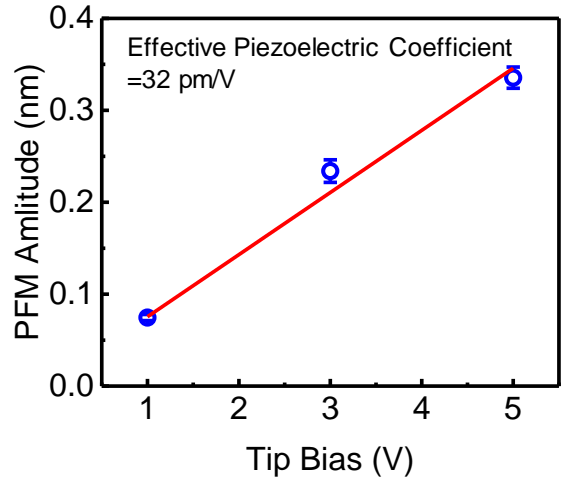

**Supplementary Figure 1.** Effective piezoelectric coefficient of a 78.7 nm thick α-In$_2$Se$_3$ flake, measured by PFM.

To extract the piezoelectric coefficient, different AC voltages are applied on the sample from the conductive AFM tip which shows a linear relationship between the mechanical deformation (PFM amplitude) and the electric field, as shown in Supplementary Figure 1. The piezoelectric coefficient ($d_{33}$) of the α-In$_2$Se$_3$ flake is 32 pm/V. It should be noticed that the $d_{33}$ obtained here is effective piezoelectric coefficient ($d_{33,eff}$), which is affected by other tensor elements from the sample and tip-sample electrostatic interaction.

Supplementary Figure 2 shows the raw data of single point DART-PFM hysteresis loop measurement measured at room temperature. Supplementary Figure 2a shows the applied voltage biases versus time. Two cycles of triangular voltage waves are applied with both on field and off field PFM measurements, as shown in Supplementary Figure 2b. Supplementary Figure 2c-e show the PFM amplitude, phase1 and phase2 signals, measured using MSM structure. Supplementary Figure 2f-h show the PFM amplitude, phase1 and phase2 signals, measured using MOS structure. A clear ferroelectric polarization switching can be seen on both phase1 and phase2 signals, at both on field and off field and for both MSM and MOS structure. The raw data itself confirms the ferroelectricity and switchable polarization of α-In$_2$Se$_3$ for FeS-FET application.

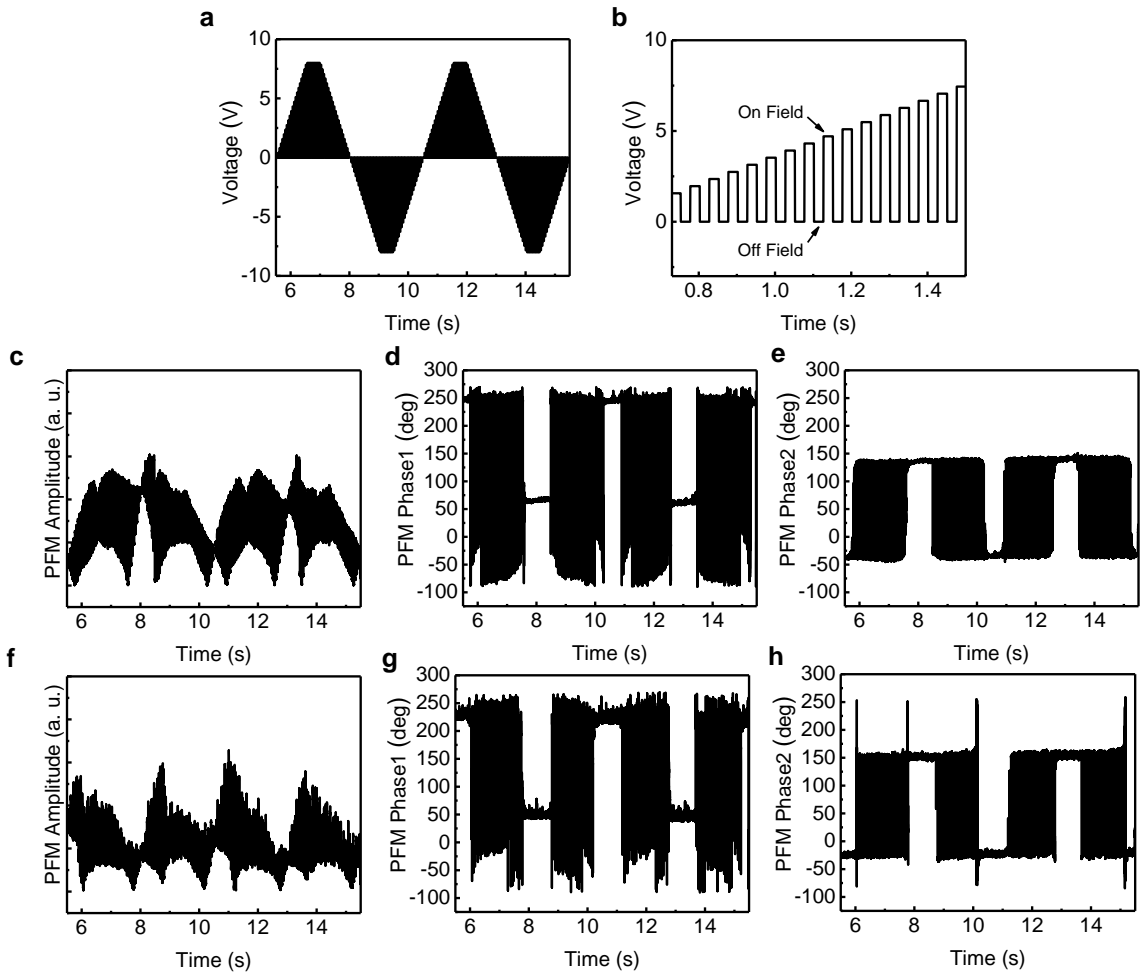

**Supplementary Figure 2.** Raw data of single point DART-PFM hysteresis loop measurements at room temperature. **a,** Bias versus time and **b,** zoom-in plot of **a**, showing on field and off field measurements. **c,** Amplitude, **d,** phase1, and **e,** phase2 results in MSM structure. **f,** Amplitude, **g,** phase1, and **h,** phase2 results in MOS structure.

## 2. Unpassivated α-In$_2$Se$_3$ FeS-FETs

Supplementary Figure 3 shows the $I_D$-$V_{GS}$ characteristics of a representative α-In$_2$Se$_3$ FeS-FET without ALD passivation, measured by double gate voltage sweep and at different $V_{DS}$. The device has a channel length ($L_{ch}$) of 1 μm, channel thickness ($T_{ch}$) of 62.2 nm. The transfer curve shows clear clockwise hysteresis loop and a large memory window over 70 V. A high on/off ratio over $10^7$ at $V_{DS}$=0.5 V between on- and off-states is also achieved. Supplementary Figure 4 investigates the impact of gate voltage sweep range and sweep time on the performance of α-In$_2$Se$_3$ FeS-FETs and are measured on a α-In$_2$Se$_3$ FeS-FET with $L_{ch}$=1 μm and $T_{ch}$=78 nm. Supplementary Figure 4a and 4b show the $I_D$-$V_{GS}$ characteristics measured at different $V_{GS}$ sweep ranges and different sweep time (sweep time controlled by the number of $V_{GS}$ step, the fastest measurement time of the whole loop is about 1 s). Supplementary Figure 4c and 4d plot the remnant $I_D$ and coercive $V_{GS}$ versus gate voltage sweep range. The gate voltage sweep range dependence suggests more polarization charge is generated by higher gate voltage. Supplementary Figure 4e and 4f show the remnant $I_D$ and coercive $V_{GS}$ versus sweep time. The $I_D$-$V_{GS}$ curve has a weak sweep speed dependence, indicating slow traps plays a minor role in device characteristics.

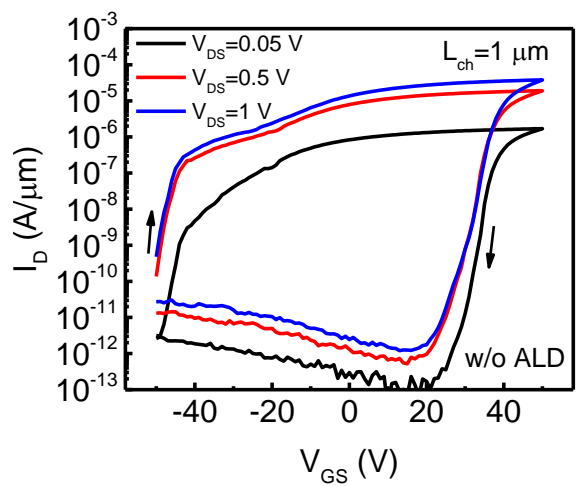

**Supplementary Figure 3**. $I_D$-$V_{GS}$ characteristics of an α-In$_2$Se$_3$ FeS-FET without ALD passivation, measured at different $V_{DS}$. The device has a channel length ($L_{ch}$) of 1 μm, channel thickness ($T_{ch}$) of 62.2 nm.

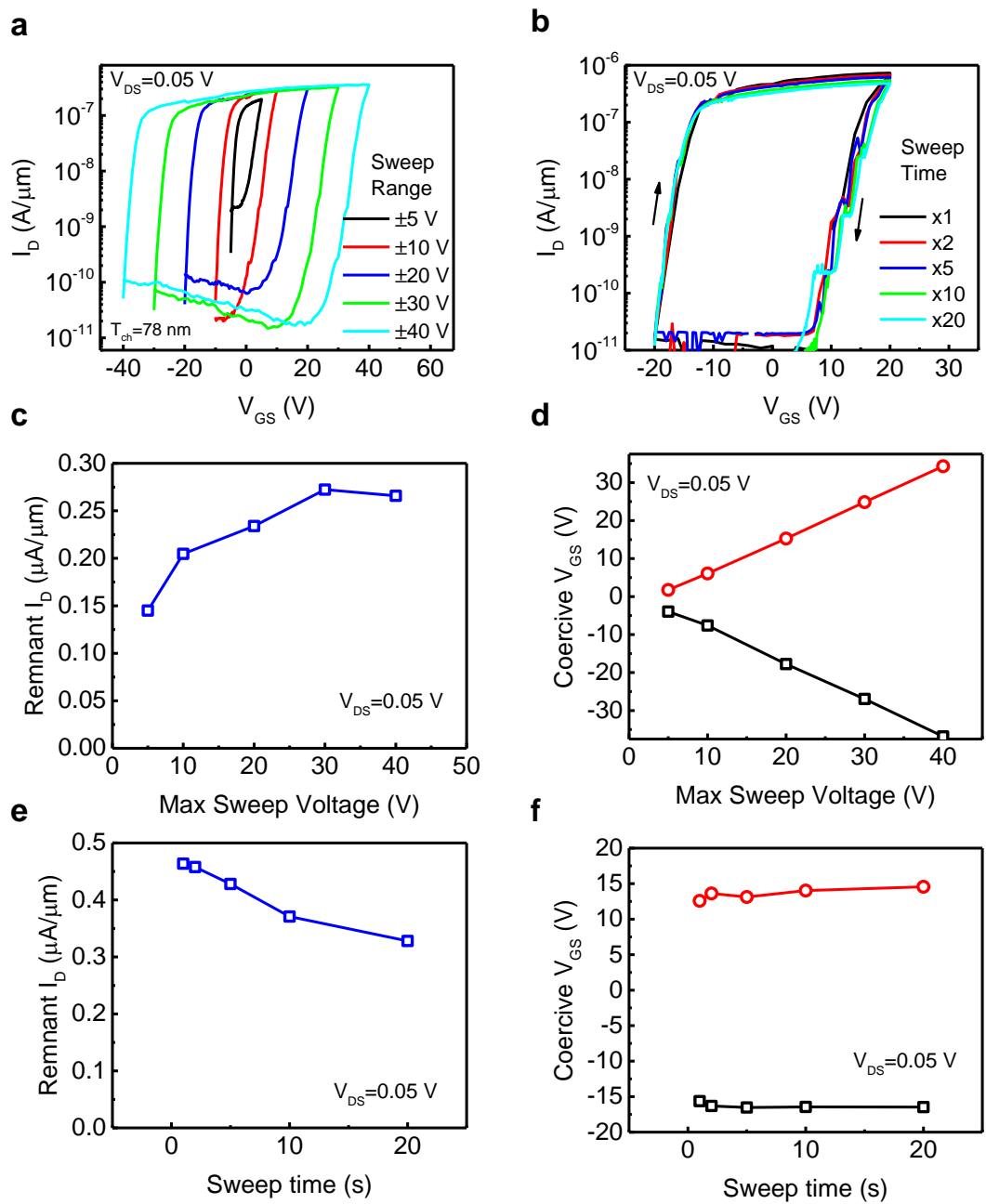

**Supplementary Figure 4**. **a,** $I_D$-$V_{GS}$ characteristics of an α-In$_2$Se$_3$ FeS-FET without ALD passivation, measured at $V_{DS}$=0.05 V and different $V_{GS}$ sweep ranges. **b,** $I_D$-$V_{GS}$ characteristics of the same α-In$_2$Se$_3$ FeS-FET, measured at $V_{DS}$=0.05 V and different sweep times. **c,** Remnant $I_D$ measured at different $V_{GS}$ sweep ranges on the same device. **d,** Coercive $V_{GS}$ ($V_{GS}$ at $I_D$= 10 nA/μm) measured at different $V_{GS}$ sweep ranges on the same device. **e,** Remnant $I_D$ measured at different $V_{GS}$ sweep times on the same device. **f,** Coercive $V_{GS}$ ($V_{GS}$ at $I_D$= 10 nA/μm) measured at different $V_{GS}$ sweep times on the same device.

## 3. Low temperature and Hall measurements

In the α-$In_2Se_3$ FeS-FETs with 90 nm $SiO_2$ as gate insulator, the hysteresis loop is clockwise. The hysteresis direction is same as the hysteresis direction induced by charge trapping. To distinguish from charge trapping induced hysteresis loop and prove the ferroelectricity, low temperature I-V and Hall measurement are performed at 80 mK. The clockwise hysteresis still exists at very low temperature measurement down to 80 mK and the hysteresis window is very similar to the room temperature results, as shown in Supplementary Figure 5a and 5b. It is well-known that charge trapping process is related with thermal activation and has strong temperature dependence. The threshold voltage shift and transistor hysteresis are reduced significantly at low temperature[1,2]. Therefore, the existence of similar hysteresis window at low temperature of 80 mK supports the existence of ferroelectricity and ferroelectric polarization switching as the origin of the hysteresis.

Supplementary Figure 5c and 5d show Hall measurements of $R_{xx}$ and $R_{xy}$ versus magnetic field (B) measured at $V_{GS}$=20 V (same device as Supplementary Figure 5a). The corresponding carrier density ($n_{2D}$) and Hall mobility ($\mu_H$) are determined to be $n_{2D}$=1.32×10$^{13}$ cm$^{-2}$ and $\mu_H$=223.4 cm$^2$ V$^{-1}$ s$^{-1}$. Supplementary Figure 5e and 5f show some additional Hall measurement data measured at $V_{GS}$=30 V and with B field measured bi-directionally (forward and reverse), suggesting no ferromagnetic property exists in this material.

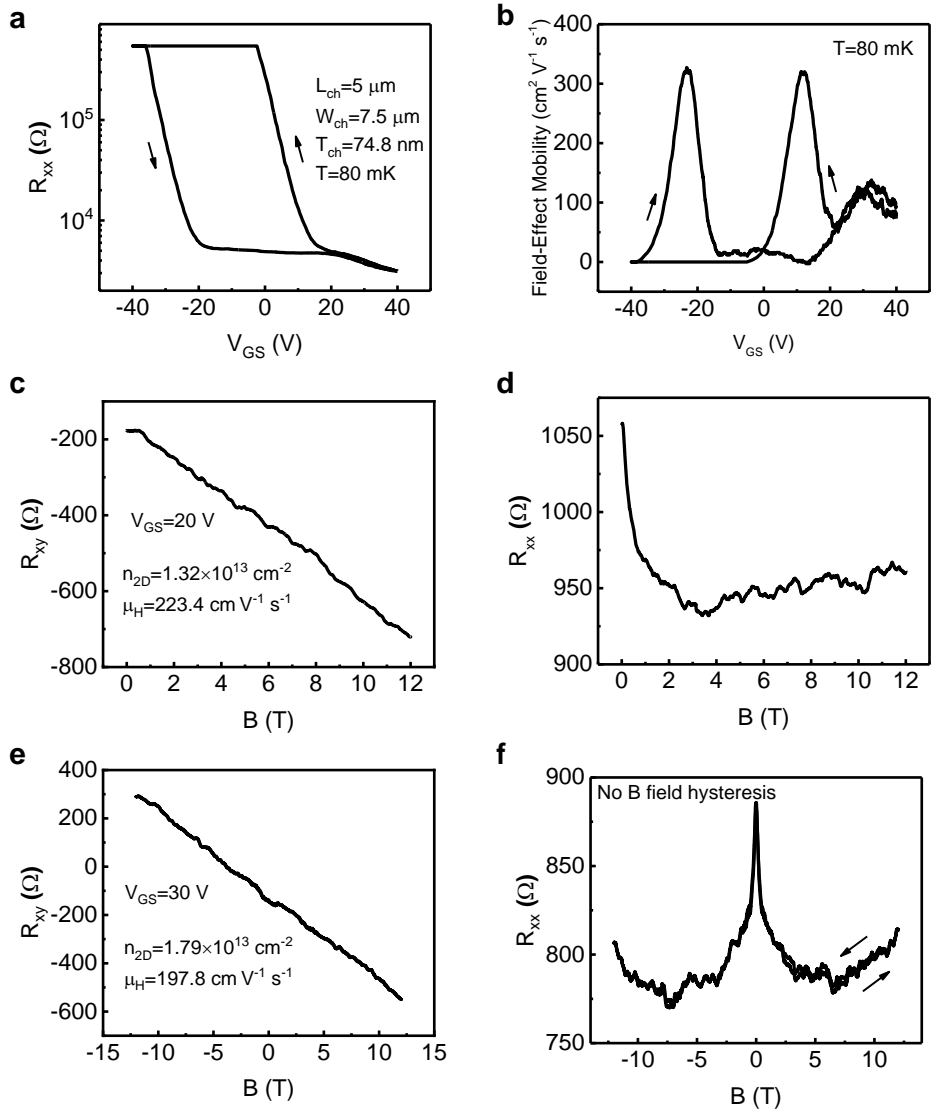

**Supplementary Figure 5. a,** $R_{xx}$-$V_{GS}$ characteristics of a α-$In_2Se_3$ FeS-FET with 90 nm $SiO_2$ as gate dielectric, 10 nm $Al_2O_3$ capping and using two terminal setup. The device has a channel length of 5 μm, channel width of 7.5 μm and channel thickness of 74.8 nm. All data in this figure are measured on the same device and at very low temperature of 80 mK. **b,** Field-effect mobility calculated from **a**. **c,** $R_{xy}$ versus B field characteristics measured at $V_{GS}$=20 V, showing a 2D electron density of $1.32\times10^{13}$ $cm^{-2}$ and a hall mobility of 223.4 $cm^2$ $V^{-1}$ $s^{-1}$. **d,** $R_{xx}$ versus B field characteristics at $V_{GS}$=20 V, showing a negative magnetoresistance with quantum oscillations due to low carrier mobility. **e,** $R_{xy}$ versus B field characteristics measured at $V_{GS}$=30 V, showing a 2D electron density of $1.79\times10^{13}$ $cm^{-2}$ and a Hall mobility of 197.8 $cm^2$ $V^{-1}$ $s^{-1}$. **f,** $R_{xx}$ versus B field characteristics measured bi-directionally at $V_{GS}$=30 V, showing a weak-localization-like negative magnetoresistance and no obvious B field hysteresis.

## 4. Channel thickness dependence in α-In$_2$Se$_3$ FeS-FETs

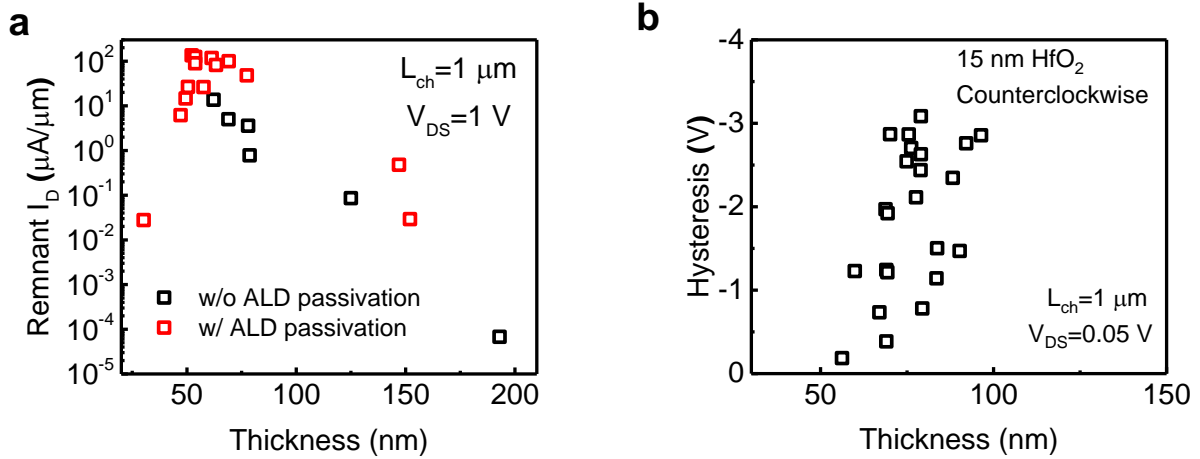

**Supplementary Figure 6**. **a,** Channel thickness dependence and comparison of remnant drain current (at $V_{GS}$=0 V and $V_{DS}$=1 V in forward sweep) versus channel thickness of α-In$_2$Se$_3$ FeS-FETs (90 nm SiO$_2$) with and without ALD Al$_2$O$_3$ passivation. Significant on-current improvement is achieved by ALD Al$_2$O$_3$ passivation. **b,** Hysteresis versus thickness for α-In$_2$Se$_3$ FeS-FETs with 15 nm HfO$_2$ as gate insulator and with 10 nm ALD Al$_2$O$_3$ passivation.

Supplementary Figure 6a shows the channel thickness dependent remnant $I_D$ of the α-In$_2$Se$_3$ FeS-FETs with $I_D$-$V_{GS}$ curve measured at $V_{DS}$=1 V. Devices with and without ALD Al$_2$O$_3$ passivation are compared. The remnant $I_D$ versus channel thickness has a peak position at around 50-70 nm, and decreases exponentially while $T_{ch}$ increases or decreases beyond this thickness range. For thicker channel, the reason of the decrease is because the maximum voltage applied (40-50 V) is not sufficiently high to trigger the ferroelectric polarization switching. For thinner channel, this might be due to the ferroelectricity in α-In$_2$Se$_3$ is getting weaker in thinner channel, but it hasn't been clearly understood yet at current stage. Therefore, the thickness dependence also suggests the ferroelectric polarization is critical to the performance and operation of the α-In$_2$Se$_3$ FeS-FETs. The maximum remnant $I_D$ at $V_{DS}$= 1V in devices with ALD passivation (135 μA/μm) is found to be significantly higher (nearly one order of magnitude) than that of devices without ALD passivation (13.7 μA/μm). The hysteresis window size is found to have a wide distribution versus thickness, as shown in Supplementary Figure 6b. The strong correlation between channel thickness and switching voltages is not obvious. This is another evidence of the competition between the bottom surface conduction and the top surface conduction. If the surface Fermi level pinning and surface passivation of the α-$In_2Se_3$ Fe-FET can be further engineered and improved, the uniformity of the hysteresis window must be improved.

## 5. Discussions on the accuracy of field-effect mobility

The field effect mobility ($\mu_{FE}$) is a defined as[3],

$$\mu_{FE} = \frac{L g_m}{W C_{ox} V_{DS}} \tag{1}$$

$$g_m = \frac{dI_D}{dV_{GS}} \tag{2}$$

The transconductance ($g_m$) is usually extracted from the transfer ($I_D$-$V_{GS}$) characteristics of the devices. It is no doubt that the $\mu_{FE}$ can always been calculated from eqns. (1) and (2). However, the accuracy of the $\mu_{FE}$ need further discussion because $\mu_{FE}$ may not be accurate to measure the intrinsic mobility ($\mu$) of the material. But it can reflect transport property of the material and the related device. There are two possible origins for the inaccurate estimation.

The first is because of the gate voltage dependence of intrinsic mobility[4]. The conductance is $\sigma = n_{ch} q \mu$, and channel carrier density $n_{ch} = C_{ox}(V_G - V_T)/q$. Thus, without considering the contact resistance, $\mu_{FE}$ can be re-write as,

$$\mu_{FE} = \frac{1}{C_{ox}} \frac{d\sigma}{dV_G} \tag{3}$$

$$\mu_{FE} = \mu + (V_G - V_T) \frac{d\mu}{dV_G} \tag{4}$$

From eqn. (4), $\mu_{FE}$ is under-estimated if $\mu$ become smaller at higher $V_G$.

The second origin is the Schottky contact resistance ($R_C$) (this time $\mu$ is assumed to be a constant for simplicity). So, we have the conductance and $\mu_{FE}$ re-write as,

$$\sigma = \frac{1}{R_c + \frac{1}{n_{ch} q \mu}} \tag{5}$$

$$\mu_{FE} = \frac{1}{(1 + n_{ch} q \mu R_c)^2} \left( \mu - n_{ch} q \mu (V_G - V_T) \frac{dR_c}{dV_G} \right) \tag{6}$$

If $R_C$ is a constant, $\mu_{FE}$ is under-estimated as shown in eqn. (7).

$$\mu_{FE} = \frac{\mu}{(1 + n_{ch} q \mu R_c)^2} \tag{7}$$

If $R_C$ is not a constant and becomes smaller at higher $V_G$, $\mu_{FE}$ can be either over- or under-estimated depends on the $R_C$ properties.

## 6. Theory and simulation of FeS-FET

Device level simulations have been conducted to investigate the clockwise and counterclockwise hysteresis in the I-V characteristics of FeS-FET. More specifically, we have performed physics based self-consistent simulation of FeS-FET devices by coupling Poisson's equation, Ginzburg-Landau equation and 2D charge equation. For simplicity, a 1D cut from the source to gate region (along $z$-axis) of FeS-FET device has been adopted in our simulation framework. A van der Waals gap is assumed between the source/drain contacts and the semiconductor channel because of the 2D layered nature of α-In$_2$Se$_3$. Now, according to the Ginzburg-Landau theory, an equation for polarization and electric field can be written as,

$$E_{FE} + K(\nabla^2 P) = \alpha P + \beta P^3 \tag{8}$$

Here, $P$ is the polarization, $E_{FE}$ is the electric-field and $K$ is the domain coupling coefficient of the ferroelectric semiconductor and all of them are defined for out-of-plane direction. Also, $\alpha$ and $\beta$ are the effective Landau coefficients that describe the dielectric stiffness of the ferroelectric semiconductor material. The P-E relation by assumption is shown in Supplementary Figure 7. Now, the total displacement ($D$) in each layer can be defined as,

$$D = \epsilon_0 \epsilon_r E_{FE} + P$$

Here, $\epsilon_0$ and $\epsilon_r$ are the vacuum and background permittivity, respectively. The Poisson's equation within the ferroelectric semiconductor layer can be written as,

$$-\epsilon_r \frac{\partial^2 \phi}{\partial z^2} = -\frac{\partial P}{\partial z} + \rho_f \tag{9}$$

$$\rho_f = q(-n + p)$$

Here, $\phi$ is the electrostatic potential, $\rho_f$ is the mobile charge density. $n$ and $p$ are the electron density and hole density, respectively and have been calculated by using the following equations.

$$n = \frac{g_C}{t_l} \times kT \times \log\left(1 + \exp\left(\frac{E_F - E_C}{kT}\right)\right); \quad g_C = \frac{2m_C^*}{\pi \hbar^2} \tag{10}$$

$$p = \frac{g_V}{t_l} \times kT \times \log\left(1 + \exp\left(\frac{E_V - E_F}{kT}\right)\right); \quad g_V = \frac{2m_V^*}{\pi\hbar^2} \tag{11}$$

Here, $g_C$ and $g_V$ are the 2D density of states of electron and hole, respectively. $m_C^*$ and $m_V^*$ are the effective mass of electron and hole, respectively. $E_C$, $E_V$ and $E_F$ are the conduction-band energy, valence-band energy and fermi-level energy, respectively. Also, $k$ is the Boltzmann constant, $\hbar$ is the reduced Plank's constant, $t_l$ is the thickness of each α-In₂Se₃ layer and $T$ is the temperature. In simulation, source electro-chemical potential has been assumed to be equal to the channel fermi-level. Finally, to get the electrostatic device characteristics we solve equation (8)-(11) self consistently to calculate polarization distribution, charge and potential profile.

From the self-consistent electrostatic simulation, we get the value of electron and hole charge density. By assuming the electron is the dominant contributor in the conduction mechanism, a drift current has been approximated by using the following equation.

$$I'_{DS} = [Q'_e + 1 \times 10^{-7} Q'_h] \times \mu \times \frac{V_{DS}}{L_{CH}}$$

$$Q_{e(h)}' = \sum_{i=1}^{n} q \times t_l n(p)_i$$

Here, $I'_{DS}$ is the current per unit area. $Q_{e(h)}'$ is the channel electrons (holes) per unit area, respectively, which have been calculated by summing up the mobile electron (hole) charges in all the α-In₂Se₃ layers.

The coupled ferroelectric and semiconducting nature of α-In₂Se₃ is critical to analyze for understanding the device operation of FeS-FET. Each of the α-In₂Se₃ layers can exhibit either a positive or negative polarization and that polarization can be switched by applying a gate voltage. However, the extent of polarization switching dependents on the amount of electric field in the α-In₂Se₃ channel and that further is dependent on the gate dielectric thickness for an applied gate voltage. To that effect, we consider two different EOT: high EOT (30 nm) and low EOT (0.5 nm) to analyze the device operations.

- High EOT devices:

First, let us assume that the applied gate voltage is highly negative (-35 V) and therefore, all the α-In$_2$Se$_3$ layers exhibit polarization down (as shown in Supplementary Figure 8b(i)). Now, with the increase in gate voltage, polarization starts to change and the bottom α-In$_2$Se$_3$ layer (at gate oxide/semiconductor interface) conduction band goes below the fermi level (Supplementary Figure 8b(i)-(iii)). Consequentially, electron density increases near the bottom α-In$_2$Se$_3$ layer and that gives rise to an increase in channel current as shown in Supplementary Figure 8b(i)-(iii). At such scenario, bottom α-In$_2$Se$_3$ layers are the most dominant participant in channel current. Interestingly, with a further increase in gate voltage, the polarization starts switching (from down to up) from the bottom layer of α-In$_2$Se$_3$ (Supplementary Figure 8b(iv)). At this point, the channel exhibits both up polarization (in first or first few bottom layers) and down polarization (in rest of the top layers) and that gives rise to a domain-wall (DW) within the channel. Note that, after the down to up polarization switching in bottom few α-In$_2$Se$_3$ layers, the minima of conduction band edge shifts from the bottom most layer to the vicinity of the DW. We refer such situation as the 'partially polarization switching' and in our considered case such phenomena happen at the gate voltage of ~30V. Now, with the decrease in gate voltage, the conduction band moves upward and consequently, current decreases rapidly. Such operation leads to clockwise hysteresis in $I_D$-$V_{GS}$ characteristics.

- Low EOT devices:

Similar to the high EOT devices, with the increase in gate voltage channel current also increases in low EOT devices (as shown in Supplementary Figure 9b(i)-(iii)). Once the bottom layer polarization is switched at a certain positive gate voltage, further increase in gate voltage will induce a layer by layer gradual polarization switching. At the same time, the α-In$_2$Se$_3$ layer corresponding to the DW moves further from the bottom to the top α-In$_2$Se$_3$ layer. At a sufficiently high and positive gate voltage, all the layer will switch to polarization up and the corresponding band diagram is being depicted in Supplementary Figure 9b(iv)-(v). In such condition, both the top and bottom layers are conducting. Now, with the decrease in gate voltage, the conduction band edge near the bottom layers will start going above the fermi-level and therefore, the electron density contribution from bottom layer will decrease (Supplementary Figure 9b(v)-(vi)). However, the top layer is still conducting. Hence, a significant amount of current can be observed even at a negative gate voltage (-3.24V). However, at a sufficiently high negative voltage, all the α-In$_2$Se$_3$ layers will switch back to down polarization and consequently, the top layers will be no more conducting. Therefore, an abrupt decrease in current can be observed (as shown in Supplementary Figure 9b(vi)-(i)). Such operation leads to a counterclockwise hysteresis in I$_{DS}$-V$_{GS}$ characteristics of FeS-FET.

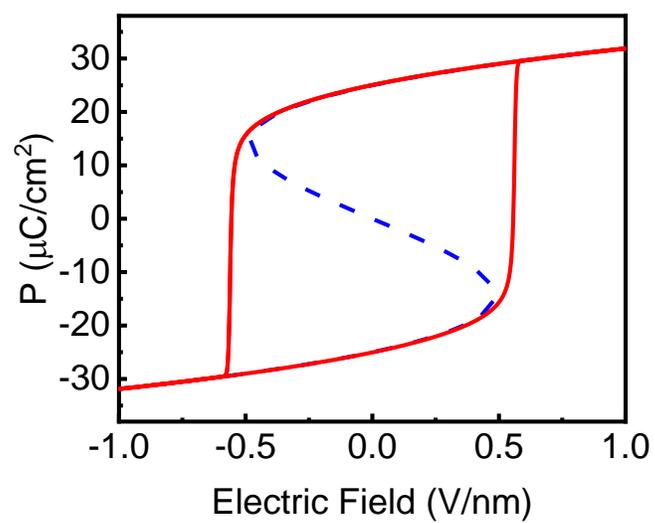

**Supplementary Figure 7.** Polarization versus electric field characteristics by assumption in the simulation.

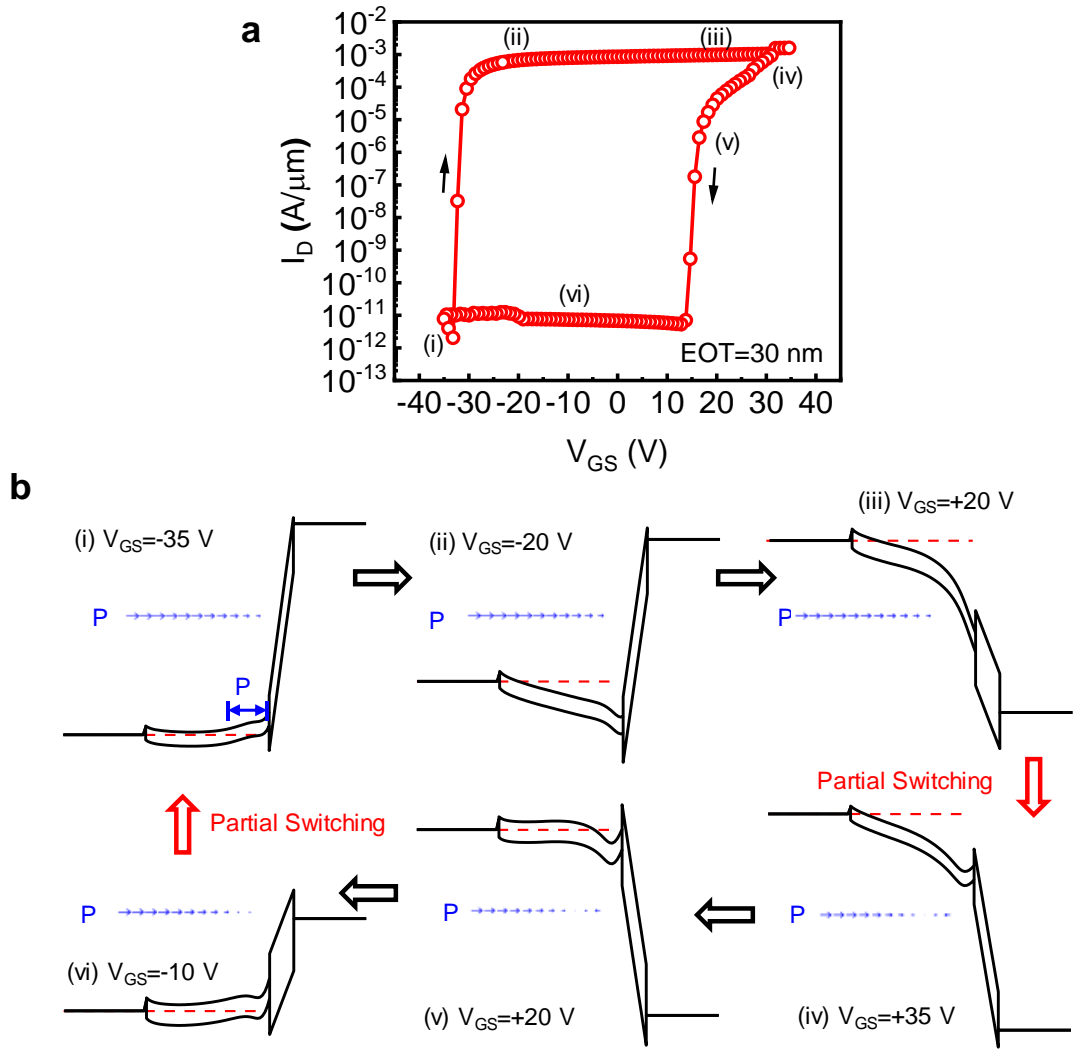

**Supplementary Figure 8. a,** Simulation of $I_D$-$V_{GS}$ characteristics of α-$In_2Se_3$ FeS-FET at EOT=30 nm. **b,** Band diagram and polarization vector map at different gate voltages during bi-directional gate voltage sweep. The direction of polarization vectors near insulator/semiconductor interface in (iv)-(vi) is opposite to those in bulk and top surface.

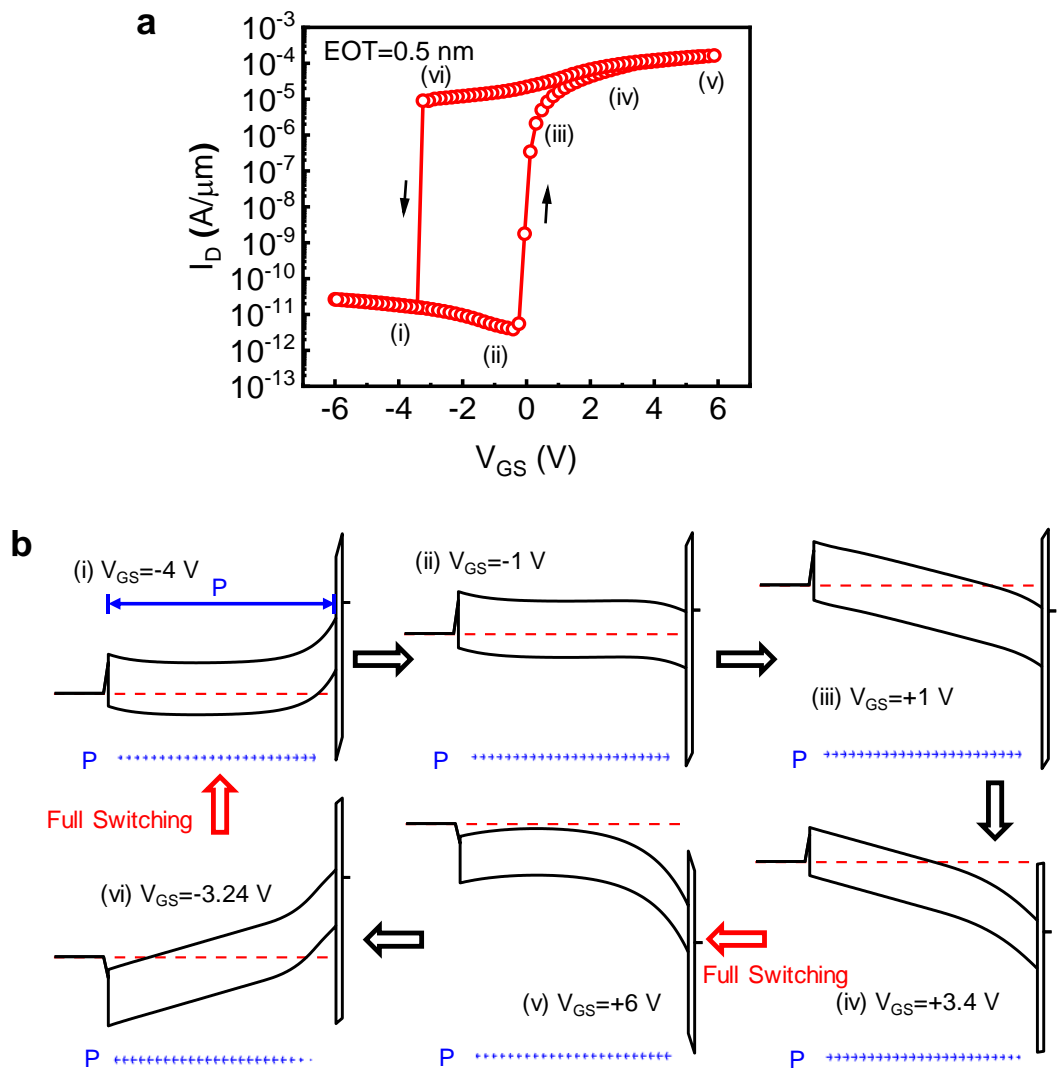

**Supplementary Figure 9. a,** Simulation of $I_D$-$V_{GS}$ characteristics of α-$In_2Se_3$ FeS-FET at EOT=0.5 nm. **b,** Band diagram and polarization vector map at different gate voltages during bi-directional gate voltage sweep.

## 7. Partial Switching of FeS-FET in Low EOT Condition

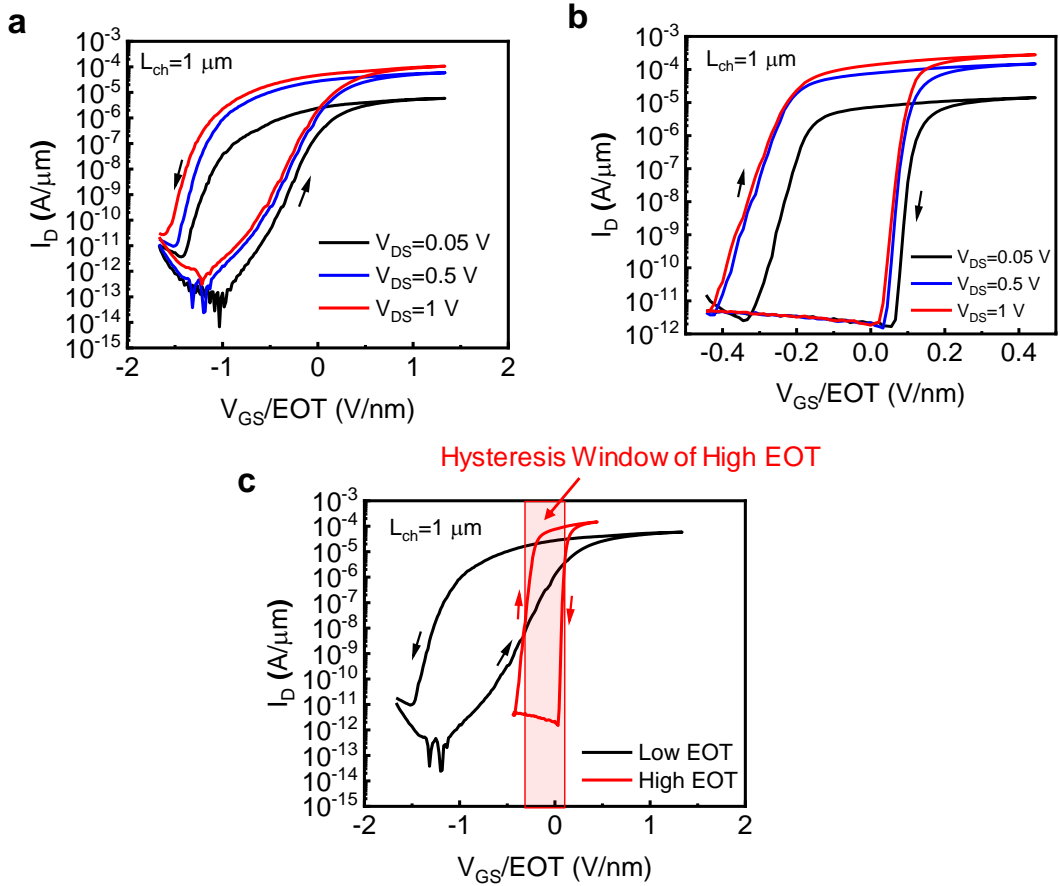

**Supplementary Figure 10**. **a,** $I_D$ as a function of $V_{GS}$/EOT for FeS-FET with low EOT. **b,** $I_D$ as a function of $V_{GS}$/EOT for FeS-FET with high EOT. **c,** Comparison of $I_D$ as a function of $V_{GS}$/EOT between low EOT and high EOT at $V_{DS}$=0.5 V.

To compare the electrical performance of α-In$_2$Se$_3$ FeS-FETs with different EOT under similar displacement field in gate oxide, we re-plot the transfer characteristics by introducing new x-axis as $V_{GS}$/EOT. Note that as displacement field is continuous (if no mobile charge), same displacement field in oxide is equivalent to same electric field in semiconductor. Supplementary Figure 10a shows $I_D$ as a function of $V_{GS}$/EOT for a FeS-FET with low EOT and Supplementary Figure 10b shows $I_D$ as a function of $V_{GS}$/EOT for a FeS-FET with high EOT. The slope in the subthreshold region is equal to SS/EOT by definition. Supplementary Figure 10c shows the comparison of $I_D$ as a function of $V_{GS}$/EOT between low EOT and high EOT at $V_{DS}$=0.5 V. It is clear that device with high EOT shows a much sharper slope (SS/EOT) in subthreshold region in this new $I_D$ versus $V_{GS}$/EOT curve. SS/EOT is much smaller at high EOT condition. A direct result is that in the hysteresis window of high EOT condition (partial switching condition), as illustrated in Supplementary Figure 10c, the device with low EOT is either at off-state in the subthreshold region or at on-state. Therefore, if applying this partial switching condition to the device with low EOT, it cannot be effectively turn-on and turn-off. In addition, to change the electric field direction inside the semiconductor, on-state to off-state or off-state to on-state transitions are necessary (to change the direction of band bending). The change of applied electric field direction is the necessary condition of ferroelectric polarization switching in the ferroelectric semiconductor. In the on-state or off-state at low EOT condition, by applying same $V_{GS}$/EOT window, the electric field in the semiconductor doesn't even change the direction (Supplementary Figure 10c), so a partial switching induced clockwise hysteresis loop is difficult to achieve for low EOT devices.

# 8. A proposal of a deep steep-slope and hysteresis-free all ferroelectric Fe²-FET

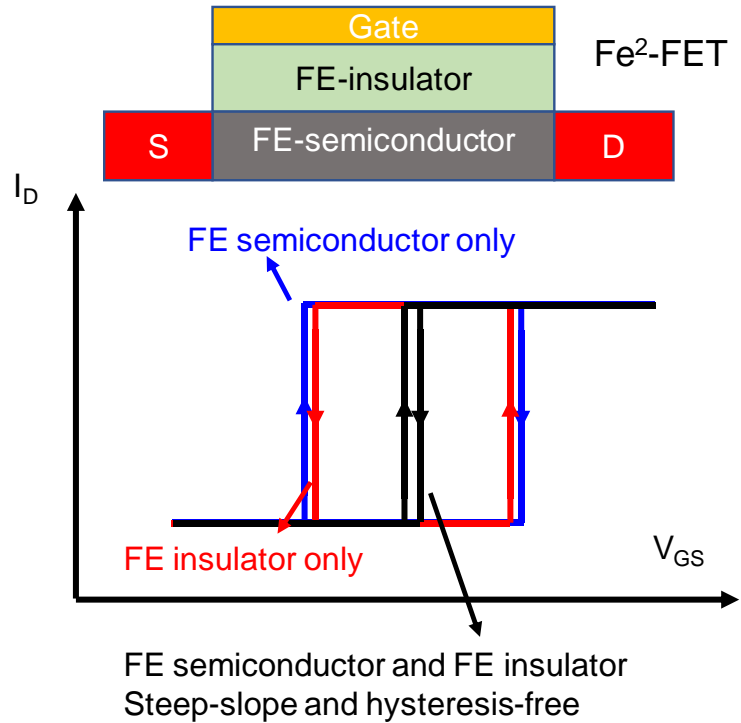

**Supplementary Figure 11**. Illustration of a deep steep-slope and hysteresis-free ferroelectric insulator and ferroelectric semiconductor all ferroelectric field-effect transistors (Fe$^2$-FET).

As another step further by using the counterclockwise hysteresis of a Fe-FET and clockwise hysteresis of a FeS-FET, we propose to integrate a ferroelectric insulator on a ferroelectric semiconductor channel, as shown in Supplementary Figure 11 as a Fe$^2$-FET, to eliminate the hysteresis loops in both Fe-FET and FeS-FET and achieve a new type of deep steep-slope and hysteresis-free transistor. The key point is to match the two ferroelectric polarization charge so that the net ferroelectric polarization charge on the FE-insulator/FE-semiconductor interface is close to zero, so that hysteresis-free operation can be achieved.